\documentclass[12pt]{iopart}

\usepackage{iopams}  
\usepackage{graphicx,subcaption}
\usepackage{rotating}
\usepackage[hidelinks]{hyperref}
\usepackage{xcolor}
\hypersetup{
    colorlinks,
    linkcolor={blue!100!black},
    citecolor={blue!100!black},
    urlcolor={blue!100!black}
}
\newcommand{\orchid}[1]{}

\begin{document}

\title[Spherical accretion onto higher-dimensional Reissner-Nordstr\"{o}m Black Hole]{Spherical accretion onto higher-dimensional Reissner-Nordstr\"{o}m Black Hole}

\author{Bibhash Das$^1$\footnote{ORCID : 0000-0001-7283-6745}, Anirban Chanda$^2$\footnote{ORCID : 0000-0002-0536-7720} and Bikash Chandra Paul$^{3}$\footnote{ORCID : 0000-0001-5675-5857} \footnotetext{Author to whom any correspondence should be addressed}}

\address{Department of Physics, University of North Bengal, Darjeeling, West Bengal, India -734 013}
\ead{$^1$rs\_bibhash@nbu.ac.in, $^2$anirbanchanda93@gmail.com, $^3$bcpaul@nbu.ac.in}

\begin{abstract}
We obtain relativistic solutions of spherically symmetric accretion by a dynamical analysis of a generalised Hamiltonian for higher-dimensional Reissner-Nordstr\"{o}m (RN) Black Hole (BH). We consider two different fluids namely, an isotropic fluid and a non-linear polytropic fluid to analyse the critical points in a higher-dimensional RN BH. The flow dynamics of the fluids are studied in different spacetime dimensions in the framework of Hamiltonian formalism. The isotropic fluid is found to have both transonic and non-transonic flow behaviour, but in the case of polytropic fluid, the flow behaviour is found to exhibit only non-transonic flow, determined by a critical point that is related to the local sound speed. The critical radius is found to change with the spacetime dimensions. Starting from the usual four dimensions it is noted that as the dimension increases the critical radius decreases, attains a minimum at a specific dimension ($D>4$) and thereafter increases again. The mass accretion rate for isotropic fluid is determined using Hamiltonian formalism. The maximum mass accretion rate for RN BH with different equations of state parameters is studied in addition to spacetime dimensions. The flow behaviour and mass accretion rate for a change in BH charge is also studied analytically. It is noted that the maximum mass accretion rate in a higher-dimensional Schwarzschild BH is the lowest, which however, increases with the increase in charge parameter in a higher-dimensional RN BH.
\end{abstract}

%
\submitto{\CQG}
%
%
%

\section{Introduction}
\label{introduction}

In astrophysics, Einstein's General Theory of Relativity (GR) is one of the most promising theories of gravity for investigating ultra-dense compact objects \cite{E1,E2,E3}. The existence of the space-time singularity or Black Hole (BH) was predicted by Schwarzschild in 1916 as an exact mathematical vacuum solution of Einstein's field equations in GR \cite{sch16a,sch16b}. The revelation of the first-ever image of a BH shadow in the centre of the M87 galaxy is a strong credibility of Einstein's GR \cite{EHT}. Another interesting prediction in GR is the accretion phenomenon in understanding the BHs. In relativistic astrophysics, accretion is a process through which matter is drawn in and absorbed by a gravitating system towards its center, which increases the mass and the angular momentum of the accreting body. It is known that accretion typically occurs when a compact object is a part of a binary system or surrounded by a disk or a cloud of gas. The matter from a companion star or the interstellar medium spirals inwards towards the compact object. Stars and planets might have formed due to accretion in some inhomogeneous regions containing gas and dust. Accretion also plays an important role in the formation and development of supermassive BHs at the center of spiral galaxies. An accumulation circle (accretion disk) also appears due to the spiralling of gas and dust clouds around the massive compact objects. However, accretion does not always increase the mass of the accreting body; it could also decrease as is observed for accretion of phantom energy \cite{jamil08,jamil09}. The presence of an event horizon, which acts as a boundary through which the infalling matter disappears in the case of accretion onto the BH, may have many implications in BH physics. For example, it gives the inner boundary condition that describes how the fluid flows and helps to avoid the uncertainties regarding the correct boundary conditions \cite{eo15}. 

Spherical accretion, also known as Bondi accretion, is a key model that explains how matter is caught and absorbed by BHs using a symmetric, non-rotating flow. The concept of spherical accretion was initially developed by Hoyle and Lyttleton \cite{hl1,hl2,hl3} for understanding stellar accretion. In 1952, Bondi investigated and extended the accretion process within the Newtonian framework, which provides a detailed mathematical framework for the accretion of gas by a gravitational body from an isotropic medium at infinity \cite{bondi52}. In 1972, Michel extended Bondi's work into the relativistic regime and obtained the accretion model, which is most acceptable in describing the accretion flow near BHs, particularly for non-rotating Schwarzchild back hole \cite{michel72}. The Bondi-type accretion flows in Schwarzchild BHs, Schwarzchild de-Sitter BHs, and Schwarzchild anti-de-Sitter BHs are found in the literature \cite{kar13,mac13a,mac13b}. Moncrief analyzed the stability of the Michel-type accretion in the subsonic region \cite{mon80}. It was shown by Jamil $et.al.$ \cite{jamil08,jamil09} and Babichev $et.al.$ \cite{bab04} that BHs may also lose mass due to the accretion of phantom energy onto it. Debnath \cite{deb15} obtained a static accretion of matter onto a general static BH with spherical symmetry using the work of Babichev and others \cite{bab04}. The radial flow of fluid is investigated in the literature, including the accretion of dark matter onto BHs \cite{gc11} and radial accretion onto cosmological BHs \cite{kar13,mac13a}, for self-gravitating objects \cite{mal99,lc14} and in the presence of perturbation theories \cite{ananda15}. Ahmad $et. \,  al.$  investigated the accretion for cyclic and heteroclinic flows near BHs obtained from $f(R)$ and $f(T)$ gravity \cite{ahmad16a,ahmad16b}. The accretion flows for polytropic fluids on Schwarzschild BHs with adiabatic index lying in the range $1 < \Gamma \leq 5/3$ have been probed in Ref. \cite{cha12}, which later extended for $\Gamma \geq 5/3$ \cite{cha16}. The spherical accretion of matter onto a charged BH in $f(T)$ gravity is reported in Ref. \cite{rod18}. In the literature  \cite{long22,bau22,feng22,rehman23,mustafa23}, recent studies on accretion onto different BHs can be found.

The success of superstring theory (SST), which is the most promising candidate for a unified theory of everything, is led to believe the existence of higher dimensions for its consistency \cite{zwi}. Although the existence of higher dimensions is not tested in the laboratory, it is still interesting to explore the physical phenomena in the framework of higher dimensions. Kaluza \cite{kaluza21} and Klein \cite{klein26} first independently introduced one extra dimension to unify electromagnetic interaction with gravity, but the approach did not work well. Eddington \cite{edi24} described astrophysical objects that are known in the usual four-dimensional spacetime, embedded in a flat higher-dimensional spacetime by considering higher dimensions. Mandelbrot \cite{man82} investigated the dimension variability issue, which described how a point object at a far way distance appears to be a three-dimensional ball at a closer distance. Emparan and Real \cite{er02} obtained relativistic solutions that describe a rotating black ring in five dimensions, indicating other topologies in higher dimensions. Cassisi $et\, al.$ \cite{cas00} investigated the stellar evolution considering the effects of higher dimensions. Several modified theories in higher dimensions have also been proposed, namely, Lovelock gravity \cite{love71,love72}, EGB gravity \cite{wil86,whe86}, etc. In Braneworld scenario \cite{mk10} a (3+1)-dimensional brane is embedded in a (4+1)-dimensional bulk. The braneworld gravity is consistent at very high enough energy and it admits a higher-dimensional geometry \cite{dl01,se19}. As the temperature decreases, there is a splitting of the spacetime where the extra dimensions are curled up. Myers and Perry \cite{mp86} extended the Schwarzschild BH solutions to $(N+1)$ dimensional spacetime. Thermodynamics of higher-dimensional spinning BHs was investigated in Ref.\cite{ag87}. Black hole solutions of higher-dimensional vacuum gravity and higher-dimensional supergravity theories are studied in Ref.\cite{er08}. Recently, Andrade $et.\, al.$ \cite{and19} constructed a higher-dimensional BH along with a new charged rotating black bar solution and studied the quasinormal modes and instabilities. Development of general parametrisation of spherically symmetric and asymptotic flat higher-dimensional BH spacetime in an arbitrary metric theory of gravity is discussed in Ref.\cite{kono20}. Frassino $et.\, al.$ \cite{fra22} used holographic braneworlds to present a higher-dimensional origin of extended BH thermodynamics. There is a spurt in research activities in the astrophysical and cosmological model building with an extension of spacetime dimensions  \cite{bs90,bcp2001,wett09}. Recently different cosmological and astrophysical missions are pouring observational data which may give the signature of extra dimensions in future. Ref.\cite{lemos24} investigate eventual deviations in the BH shadow radius using the observational data of the Sgr A* shadow released by ETH to dig out physics coming from extra dimensions. More fascinating details on higher-dimensional BHs can be found in the literature  \cite{tan63,em00,rod18,sch20,ast20,dadhich22,van23,lobos24}. \\

The BH accretion in higher dimensions is an interesting topic of research.  John $et.\, al.$ \cite{john13} investigated the steady-state spherically symmetric accretion of relativistic fluids onto a higher-dimensional Schwarzschild BH with a polytropic equation of state (EoS). Polytropic fluid accretion onto a higher-dimensional RN BH is discussed in Ref.\cite{sha16}. Liu $et.\, al.$ \cite{liu24} probed the characteristics like energy flux, radiation temperature and spectral cutoff frequency of the accelerating disks surrounding a braneworld BH.  Accretion onto higher-dimensional BHs is also discussed in Ref.\cite{sha11,saa20}. \\

 Bondi \cite{bondi52} studied the accretion of barotropic fluid onto BHs in the Newtonian framework. Michel \cite{michel72}, and others \cite{john13,sha16} extended this work in the framework of general relativity by investigating a spherically symmetric flow onto a Schwarzschild BH. Recently, a dynamical system, namely, the Hamiltonian approach in a non-Newtonian framework, has been employed to study spherical accretion onto BHs in a sophisticated theoretical framework that describes the dynamics of particles and fields in the curved spacetime surrounding a BH \cite{ahmad16a,ahmad16b,cha12,cha16}. This method is particularly useful because it provides a clear and systematic way to derive the equations of motion for infalling matter radially into a BH, utilising the principles of Hamiltonian mechanics in the context of GR.  Moreover, with the Hamiltonian dynamical analysis we can obtain the type of equilibrium at the critical points of the dynamical system. In the present paper, we use a Hamiltonian dynamics method to investigate the flow behaviour and the accretion rate of accreting fluid assuming two different EoS onto a higher-dimensional Reissner–Nordstr\"{o}m (RN) BH \cite{xie22}. Here, isothermal and polytropic fluids are taken for mathematical modelling to analyse the behaviour of spherical accretion in a higher-dimensional BH. The fluids simplify the mathematical equations describing the disk's behaviour by assuming constant temperature throughout. Consequently, one can focus on basic physics, such as gas flow dynamics, charge variation, and dimensional effects. Comparing the flows in the theoretical models with the observations of the known fluids in the laboratory, we derive knowledge for further understanding of accretion with the isothermal and polytropic fluid flow at a constant temperature. This leads to a better comprehension of the physical mechanisms that operate in the accretion disks and the viability of theoretical models. Sonic points or critical points correspond to the condition that the velocity of the fluid equals the local speed of sound and are determined by analysing the Hamiltonian dynamical system. The sonic points further help us to critically analyse the flow of fluid near the event horizon of the BH. In the case of isothermal fluid ($p = \omega \rho$), the accretion is studied for different values of the state parameter $\omega$, whereas the polytropic fluid ($p = \kappa \varrho^{\Gamma}$) is studied for a specific value of the adiabatic index $\Gamma$. We also explore the accretion of matter onto an RN BH with a variation of charge.

The paper is organised as follows: Section \ref{sec:ge},  the general equations and conservation laws for spherical accretion are derived. The speed of sound at the sonic (critical) points is determined using the Hamiltonian dynamical system in Section \ref{sec:3}. In Section \ref{sec:4}, we analysed and discussed the characteristics of accretion flow near a higher-dimensional RN BH for different types of fluids such as ultra-stiff fluid (USF), ultra-relativistic fluid (URF), radiation fluid (RF) and sub-relativistic fluid (SRF). The polytropic test fluid accretion is considered in Section \ref{sec:5} to find the accretion types near the RN BH. In section \ref{sec:massrate} we discussed BHs mass accretion rate for isotropic fluid. In Section \ref{sec:6}, we discussed and compared the effects of charge variation in accretion for different dimensions. Finally, in Section \ref{sec:7}, we briefly summarise the results obtained in this case.


\section{Fundamental equations for spherical accretion flow}
\label{sec:ge}
We consider a static spherically symmetric D-dimensional Reissner-Nordstr\"{o}m (RN) metric, described by the line element,
\begin{equation}
	\label{lineelement}
	dS^2 = -f(r) dt^2 + \frac{dr^2}{f(r)} + r^2 d \Omega^2_{D-2},
\end{equation}
where, $D$ is the spacetime dimension and $\Omega_{D-2}$ is the area of a unit $(D-2)$-dimensional sphere
\begin{equation}
	\Omega_{D-2} = \frac{2\pi^{\frac{D-1}{2}}}{\Gamma(\frac{D-1}{2})}, \nonumber
\end{equation}
and $d \Omega^2_{D-2}$ is the line element of the sphere, \textit{i.e.},
\begin{equation}
	d \Omega^2_{D-2} = d\theta^2_1 + \prod_{\mu=1}^{D-3} \sin^2\theta_{\mu}d\theta^2_{D-2}. \nonumber
\end{equation}
 We have used common relativistic notation,  geometric units $G=c=1$ and the metric signature $(-,+,+,+)$. The function $f(r)$ in higher-dimensional  RN metric is given by
\begin{equation}
	f(r) = 1 - \frac{16 \pi M r^{3-D}}{(D-2)\Omega_{D-2}} + \frac{32 \pi^2 q^2 r^{2(3-D)}}{(D-2)(D-3)\Omega^2_{D-2}},
\end{equation}
where $M$ and $q$ are the mass and charge, respectively. The roots of $f(r) =0$ will give us the location of the  horizon for the $D$-dimensional RN BH \cite{xie22}, which is
\begin{equation}
	\label{ehradius}
	r_{\pm} = \Bigg[  \frac{4\pi}{\Omega^2_{D-2}} \Bigg( \frac{2M}{D-2} \pm \sqrt{\frac{4M^2}{(D-2)^2} - \frac{2q^2}{(D-2)(D-3)}} \Bigg) \Bigg]^{\frac{1}{D-3}},
\end{equation}
 where, $r_{+} \equiv r_h$ is the event horizon and $r_{-} \equiv r_{ch}$ is the Cauchy horizon. We investigate spherical accretion onto a higher-dimensional RN BH, which typically examines the fluid dynamics around the event horizon of the BH. To begin with, we consider perfect fluid to analyse the steady-state flow of fluid in radial direction onto a higher-dimensional Reissner-Nordstr\"{o}m BH. We consider two basic laws that characterise the spherical accretion of a perfect isotropic fluid, \textit{i.e.} the particle conservation law and the energy-momentum conservation law. In the first case, the conservation of particles in the process is described by the conservation law :
\begin{equation}
	\label{particleconservation}
	\nabla_{\mu} J^{\mu} = 0,
\end{equation}
where, the baryon number flux is, $J^{\mu} = \varrho \,u^{\mu}$, $\varrho$ represents the baryon number density, and $u^{\mu} = \frac{dx^{\mu}}{d\tau}$ represents D-velocity of the fluid. On the other hand, the energy-momentum conservation law is given by 
\begin{equation}
	\label{energyconservation}
	\nabla_{\mu} T^{\mu \nu} = 0,
\end{equation}
The energy-momentum tensor for a perfect fluid is $T^{\mu \nu} = (\rho + p)\,u^{\mu} u^{\nu} + p \,g^{\mu \nu}$, where $\rho$ and $p$ are the energy density and pressure, respectively. We assume a steady flow of the fluid in the equatorial plane ($\theta = \frac{\pi}{2}$), yielding non-zero components of the D-velocity of the flow as follows: $u^t = \frac{dt}{d\tau}$ and $u^r = \frac{dr}{d\tau} = u$. We use  the normalization condition $u^{\mu}u_{\mu} = -1$, which gives 
\begin{equation}
	(u^t)^2 = \frac{u^2 + f(r)}{(f(r))^2}.
\end{equation}
On the equatorial plane ($\theta = \frac{\pi}{2}$), the continuity equation (\ref{particleconservation}) becomes
\begin{equation}
\label{de1}
	\frac{1}{r^{D-2}}\frac{d}{dr}\Big( r^{D-2} \varrho \, u \Big) = 0.
\end{equation}
On integration, it yields
\begin{equation}
	\label{maineq1}
	r^{D-2} \varrho \, u = C_1.
\end{equation}
where $C_1$ is an integration constant. Again applying the first law of thermodynamics for perfect fluid we  obtain the following
\begin{equation}
	\label{FLT}
	dp = \varrho(dh - T \,ds), \;\;\; d\rho = h \,d\varrho + \varrho \,T ds,
\end{equation} 
where $T$ is the temperature, $s$ is the specific entropy (entropy per particle), and $h = \frac{\rho + p}{\varrho}$ is the specific enthalpy (enthalpy per particle). The fluid motion is considered here to be radial, stationary (independent of time), and isentropic ($s =$ const.). Consequently, the above equation reduces to
\begin{equation}
	\label{isentropic}
	dp = \varrho \,dh, \;\;\; d\rho = h\, d\varrho.
\end{equation}
Since the flow of the fluid is smooth, from equation (\ref{energyconservation}) we obtain
\begin{equation}
	\varrho \,u^{\mu}\,\nabla_{\mu}(h\, u^{\nu}) + g^{\mu \nu}\,\partial_{\mu}p = 0.
\end{equation}
For an isentropic fluid along a streamlined flow, we get
\begin{equation}
	u^{\mu}\,\nabla_{\mu}(hu^{\nu}) + \partial_{\nu} h = 0.
\end{equation}
Taking the zeroth component of the above equation, we get
\begin{equation}
\label{de2}
	\partial_r(h\,u_t) = 0.
\end{equation}
On integration, we obtain 
\begin{equation}
	\label{maineq2}
	h \Big[ f(r) + u^2 \Big]^{1/2} = C_2,
\end{equation}
where $C_2$ is the integration constant. Equations (\ref{maineq1}) and (\ref{maineq2}) are the two important equations that are used to analyse the flow of a perfect fluid in the background of a higher-dimensional RN BH.

The speed of sound ($a$) is $a^2 = (\frac{\partial p}{\partial \rho})_s$. Since entropy ($s$) is a  constant, the speed of sound becomes $a^2 = \frac{d p}{d \rho}$. Using equation (\ref{isentropic}), we obtain
\begin{equation}
	\label{sos}
	a^2 = \frac{\varrho \,dh}{h\,d\varrho} \; \Rightarrow \; \frac{dh}{h} = a^2 \frac{d\varrho}{\varrho}.
\end{equation}
At the equatorial plane, the particle speed is defined by $v \equiv \frac{dr}{f(r)dt}$ which can be rewritten to 
\begin{equation}
	\label{vsquared}
 v^2 = \Bigg( \frac{u}{f(r)u^t} \Bigg)^2 = \frac{u^2}{u_t^2} = \frac{u^2}{f(r) + u^2}.
\end{equation}
where 
\begin{equation}
	\label{u}
	u^2 = \frac{v^2 f(r)}{1-v^2} \;\;\; \& \;\;\; u_t^2 = \frac{f(r)}{1 - v^2},
\end{equation}
and therefore, one can determine the integration constant in equation (\ref{maineq1}), which yields
\begin{equation}
	\label{constant1}
	C_1^2 = \frac{\varrho^2\,v^2\,r^{2(D-2)}f(r)}{1 - v^2}.
\end{equation}


\section{Hamiltonian systems for Accretion and Sonic Points }
\label{sec:3}
In the above, we have derived two integrals with the integration constants ($C_1$,$C_2$), making
use of equations (\ref{maineq1}) and (\ref{maineq2}). It is important to mention that any one of the integrals or both of them can be employed to construct a Hamiltonian system $\mathcal{H}$ for understanding the fluid flow. Thus, the Hamiltonian system is a function of two variables, namely, $r$ and $v$, and we denote  $\mathcal{H}$ as the  square of LHS of the equation (\ref{maineq2}), which can be expressed as
\begin{equation}
 \mathcal{H}(r,v)	 = h^2 (f(r) + u^2).
\end{equation}
Using equation (\ref{u}), we obtain the Hamiltonian equation to study the dynamical system of fluids in the framework of BHs:
\begin{equation}
	\label{hamiltonian}
	\mathcal{H}(r,v) = \frac{h(r,v)^2 \,f(r)}{1 - v^2}.
\end{equation}
From now on, the partial derivatives will be denoted by $\frac{\partial f(r)  }{\partial x} = f(r)_{,x}$.

\subsection{Critical analysis at the Sonic points}
 In fluid dynamics, a sonic point or a critical point (CP) refers to a crucial location where the fluid's velocity becomes equal to the local speed of sound. The fluid flow that passes through a sonic point is called a transonic flow and possesses a velocity equal to the local sound speed. If the velocity of the flow is lower than the local sound speed it is called a subsonic flow, and if the velocity is greater than the local sound speed it is called a supersonic flow. At a sonic point, the flow transitions from subsonic to supersonic. Equation (\ref{hamiltonian}) can be employed to derive the sonic points of the dynamical system. With given $\mathcal{H}$ in (\ref{hamiltonian}), the dynamical system becomes
\begin{equation}
	\label{HS}
	\dot{r} = \frac{\partial \mathcal{H}}{\partial v} \;\;\;\; ; \;\;\;\;  \dot{v} = - \frac{\partial \mathcal{H}}{\partial r},
\end{equation}
where the dot represents the derivative with respect to time ($t$). Evaluating the RHS of the above equation, we get
\begin{equation}
	\frac{\partial \mathcal{H}}{\partial v} = \frac{2v\,h^2\,f(r)}{(1-v^2)^2} \Bigg[ 1+\frac{(1-v^2)}{v} \,(\ln h)_{,v} \Bigg], \nonumber
\end{equation}
\begin{equation}
	\frac{\partial \mathcal{H}}{\partial r} = \frac{h^2}{(1-v^2)} \Bigg[ f'(r) + 2f(r)\, (\ln h)_{,r} \Bigg].
\end{equation}
Equation (\ref{sos}) yields
\begin{equation}
	(\ln h)_{,v} = a^2 \,(\ln \varrho)_{,v} \;\;\;\; \mbox{and}\;\;\;\;	(\ln h)_{,r} = a^2 \,(\ln \varrho)_{,r}.
\end{equation}
We note the following:\\
(i) When $r$ is kept constant, equation (\ref{constant1}) leads to  $ \frac{\varrho v}{\sqrt{1-v^2}} =$ constant, which upon derivative w.r.t. $v$ will get us
\begin{equation}
	(\ln \varrho)_{,v} = -\frac{1}{v\,(1-v^2)} \;\; \Rightarrow\;\; (\ln h)_{,v} = - \frac{a^2}{v\,(1-v^2)};
\end{equation}
(ii) When $v$ is kept constant in equation (\ref{constant1}) we get $r^{D-2}\varrho\sqrt{f(r)} = $ constant. and its  derivative  w.r.t. $r$ yields
 \begin{equation*}
     (\ln \varrho)_{,r} = -\frac{2(D-2)+ r\, (\ln f(r))_{,r}}{2r} \nonumber
 \end{equation*}
   \begin{equation}
 \Rightarrow\; (\ln h)_{,r} = - \frac{a^2 \Big[2(D-2)+ r\, (\ln f(r))_{,r}\Big]}{2r} .
\end{equation}
Therefore, the autonomous system of the differential equations in equation (\ref{HS}) can be presented as
\begin{equation}
    \label{HS1}
	\dot{r} = \frac{2h^2\,f(r)}{v\,(1-v^2)^2} (v^2 - a^2),  
\end{equation}
\begin{equation}
     \label{HS2}
	\dot{v} = -\frac{h^2}{r\,(1-v^2)} \Big[ r\,f(r)_{,r} (1-a^2) - 2(D-2)\,a^2f(r) \Big].
\end{equation}
The sonic points can be obtained by equating the RHS of the equations (\ref{HS1}) and (\ref{HS2}) to zero, which leads to 
\begin{equation}
	\label{CP1}
	v_c^2 = a_c^2 \;\;\;\;\;  \mbox{and}  \;\;\;\;\; r_c\,(1-a_c^2)\,f(r_c)_{,r_c} = 2(D-2)\,a_c^2 f(r_c).
\end{equation}
The subscript letter $c$ in the above represents the quantities values at the sonic or critical points. Using the second equation on equation (\ref{CP1}), the expression for the speed of sound at the sonic point ($a_c$) can be obtained as
\begin{equation}
	\label{sosCP}
	a_c^2 = \frac{r_c\,f(r_c)_{,r_c}}{r_c\,f(r_c)_{,r_c} + 2(D-2)\,f(r_c)}.
\end{equation}
Now using equation (\ref{CP1}) in equation (\ref{constant1}) we determine
\begin{equation}
	\label{constant1CP}
	C_1^2 = \frac{r_c^{(2D-3)}\,\varrho_c^2\,f(r_c)_{,r_c}}{2(D-2)}.
\end{equation}
Comparing equations (\ref{constant1}) and (\ref{constant1CP}), we obtain the following
\begin{equation}
	\label{numberdensityratio}
	\Big( \frac{\varrho}{\varrho_c} \Big)^2 = \frac{r_c^{(2D-3)}\,f(r_c)_{,r_c}}{2(D-2)}\; \frac{1-v^2}{r^{2(D-2)}\,v^2\,f(r)}.
\end{equation}
The above equation permits two types of flow of the fluids through the critical point near the BH horizon. In the first case, the fluid speed $v$ vanishes, and in the second case, the fluid speed approaches the speed of light in such a way that the ratio $\frac{(1-v^2)}{f(r)}$ attains an infinite value. The number density $\varrho$ diverges on the horizon in the former case regardless of the expression of the metric potential $f(r)$.


 \subsection{Dynamical analysis}
 We now have the autonomous system of differential equations, from which we analyze the types of critical points for the dynamical system by computing a Jacobian matrix  (using equations (\ref{HS1}) and (\ref{HS2})). Firstly, we linearised the dynamical system using Taylor's expansion of equations (\ref{HS1}) and (\ref{HS2}) around the critical points which are given by
\begin{equation}
    \left(
        \begin{array}{c}
            \delta \dot{r} \\
            \delta \dot{v}
        \end{array} \right) = \mathcal{J} \left(
        \begin{array}{ccc}
            \delta r \\
            \delta v
        \end{array} \right),
\end{equation}
where, $\delta r$ and $\delta v$ are the small perturbations of $r$ and $v$ around the critical points, and $\mathcal{J}$ is the Jacobian matrix of the dynamical system at the critical point ($r_c,\,v_c$). The Jacobian matrix is defined as
\begin{equation}
    \mathcal{J} =  \Bigg(
        \begin{array}{cc}
            \frac{\partial g_1}{\partial r} & \frac{\partial g_1}{\partial v} \\
            \frac{\partial g_2}{\partial r} & \frac{\partial g_2}{\partial v} 
        \end{array} \Bigg)_{(r_c,v_c)}
\end{equation}
where, $g_1(r,v) \equiv \dot{r}$ and $g_2(r,v) \equiv \dot{v}$. Now, depending upon the eigenvalues ($\lambda_1,\,\lambda_2$) of the Jacobian matrix, the types of the critical points can be identified as follows:
\begin{itemize}
    \item Node: $\lambda_1,\,\lambda_2$ are real numbers and of the same sign                      ($\lambda_1 \cdot \lambda_2 >0$),
    \item Saddle: $\lambda_1,\,\lambda_2$ are real numbers and non-zero of the                         opposite sign ($\lambda_1 \cdot \lambda_2 <0$),
    \item Focus: $\lambda_1,\,\lambda_2$ are complex numbers, the real parts are equal                 and non-zero (\it{Re} $\lambda_1 =$ \it{Re} $\lambda_2 \neq 0$),
    \item Center: $\lambda_1,\,\lambda_2$ are purely imaginary numbers                                 (\it{Re} $\lambda_1 =$ \it{Re} $\lambda_2 = 0$).
\end{itemize}


\section{Applications to Isothermal fluid accretion}
\label{sec:4}
\label{isothermal}
Isothermal fluids are referred to as those fluids that flow at a constant temperature so that the sound speed remains constant throughout the accretion process. Due to the very fast speed of the fluid, the accretion process shifts the dynamical system to an adiabatic system. The equation of state (EoS) for this kind of fluid is given by  $p = \omega \rho$, where $\omega$ is the state parameter ($0< \omega \leq 1$). In general, the adiabatic sound speed is  $a^2 = \frac{dp}{d\rho}$. Therefore, using the EoS and defined sound speed, we obtain $a^2 = \omega$.  Using the first law of thermodynamics in equation (\ref{FLT}), we get
\begin{equation}
	\frac{d\rho}{d\varrho} = \frac{\rho + p}{\varrho} = h.
\end{equation}
Integrating the above equation from the sonic point to a point inside the fluid, we get
\begin{equation}
	\frac{\varrho}{\varrho_c} = \exp \Bigg( \int^{\rho}_{\rho_c} \frac{d\rho^{'}}{\rho^{'} + p(\rho^{'})} \Bigg).
\end{equation}
Using EoS $p = \omega \rho$, the above equation reduces to
\begin{equation}
    \label{nratio}
	\frac{\varrho}{\varrho_c} = \Bigg( \frac{\rho}{\rho_c} \Bigg)^{\frac{1}{1+\omega}}.
\end{equation}
Using equation (\ref{nratio}) we calculate the enthalpy $(h)$, which is given by
\begin{equation}
\label{enthalpy1}
	h = \frac{(1+\omega)\, \rho_c}{\varrho_c} \Bigg( \frac{\varrho}{\varrho_c} \Bigg)^{\omega}.
\end{equation}
Now, using equation (\ref{numberdensityratio}) in equation (\ref{enthalpy1}), we get
\begin{equation}
	h^2 \propto \Bigg( \frac{1 - v^2}{r^{2(D-2)}\,v^2\,f(r)} \Bigg)^{\omega}.
\end{equation}
Using $h^2$ in equation (\ref{hamiltonian}) once again, we obtain
\begin{equation}
	\label{finalhamiltonian}
	\mathcal{H} = \frac{f(r)^{1-\omega}}{r^{2\omega(D-2)}\,v^{2\omega} \,(1-v^2)^{1-\omega}}.
\end{equation}
The equations (\ref{u}) and (\ref{sosCP}), are  employed to derive another set of equations that determines the radial velocity at the critical point, which is given by 
\begin{equation}
     \label{uc1} 
	(i)\ \ u_c^2 = \frac{r_c \,f(r_c)_{,r_c}}{2(D-2)},
\end{equation}
\begin{equation}
    \label{uc2}
	(ii)\ \ u_c^2 = \omega \Bigg(  f(r_c) + \frac{r_c \,f(r_c)_{,r_c}}{2(D-2)} \Bigg). 
\end{equation}
The new Hamiltonian and the critical radial velocity at the sonic points are determined by equations (\ref{finalhamiltonian}), (\ref{uc1}), and (\ref{uc2}), respectively. We can study the flow of accretion for a given type of fluid determined by its state parameter value $\omega$. We analyze the flow of fluids taking the following values of $\omega$: $\omega = 1$, $\omega = \frac{1}{2}$, $\omega = \frac{1}{3}$, and $\omega = \frac{1}{4}$ which are referred to as ultra-stiff fluid (USF), ultra-relativistic fluid (URF), radiation fluid (RF), and sub-relativistic fluid (SRF), respectively, in the next sections.

\begin{figure*}[h]
\centering
\begin{tabular}{cc}
  \includegraphics[width=0.45\textwidth,height= 0.35\textwidth]{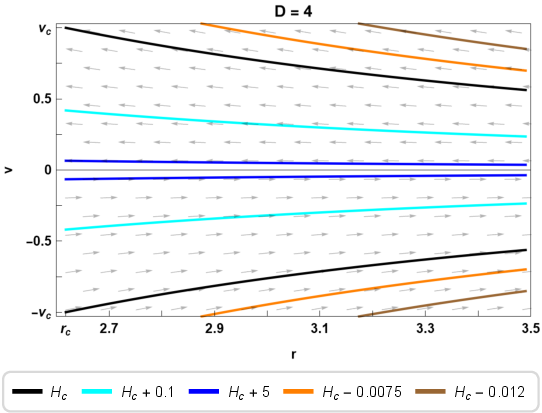} &   \includegraphics[width=0.45\textwidth,height= 0.35\textwidth]{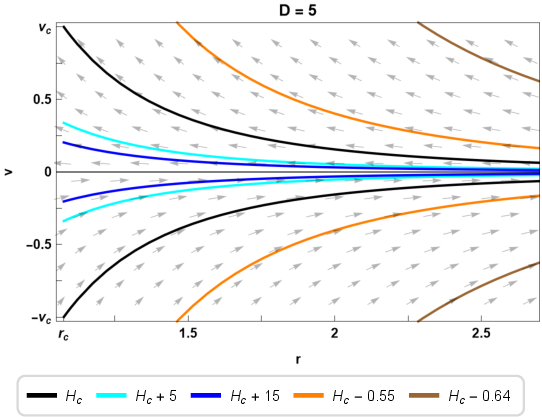}\\[6pt]
  \includegraphics[width=0.45\textwidth,height= 0.35\textwidth]{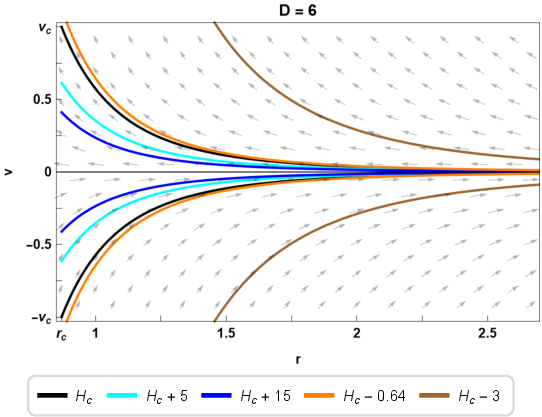} &   \includegraphics[width=0.45\textwidth,height= 0.35\textwidth]{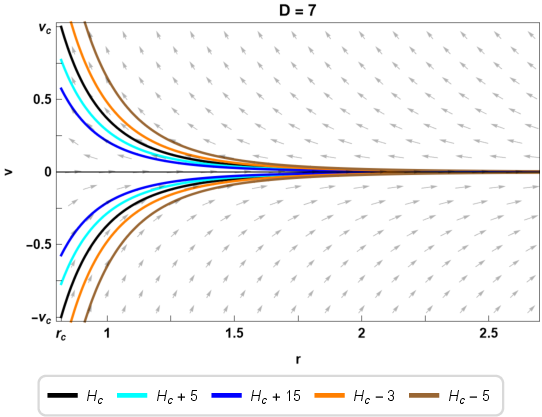}\\[6pt]
\end{tabular}
\caption{Contour plots of  Hamiltonian $\mathcal{H}$ with different dimensional BHs for the parameters $\omega = 1$, $M = 1.5$ and $q = 1$. The black curve corresponds to the critical Hamiltonian $\mathcal{H}_c$. The Blue and Cyan curves correspond to $\mathcal{H} > \mathcal{H}_c $, and the Orange and Brown curves represent $\mathcal{H} < \mathcal{H}_c$.  } 
 	\label{fig:usf}
\end{figure*}

\subsection{Case I : Ultra-Stiff Fluid ($\omega = 1$)}
The ultra-stiff fluids are described by the EoS $p = \rho$, \textit{i.e.}, where the state parameter is  $\omega=1$. Using equations (\ref{uc1}), and (\ref{uc2}), we determine the critical radius ($r_c$) for the equation  $f(r_c) = 0$. Thus, in this case  the critical radius ($r_c$) represents the event horizon radius ($r_h$), {\it i.e.},  $r_c = r_h$. The Hamiltonian, in this case, is given by
\begin{equation}
	\label{H1}
	\mathcal{H} = \Bigg( \frac{1}{v \,r^{D-2}} \Bigg)^2.
\end{equation}
 
\begin{table}[h] 
\centering
\caption{\label{tab:usf}}
Values of $r_c$, $v_c$, and $H_c$ at the sonic point for ultra-stiff fluid ($\omega  = 1$) for even and odd dimensional BHs with suitable parameters $M = 1.5$ and $q = 1$.
\begin{tabular}{c c c c c c}
        \br
	D	&	$r_h$      	&	$r_c$    	&	$v_c$	&	$H_c$      & Equilibrium point type	  \\
	\mr
	4	&	2.618030	&	2.618030	&	  1		&	0.0212862 & Center \\
	5	&	1.075370	&	1.075370	&	  1		&	0.64663   & Focus   \\
	6	&	0.869997	&	0.869997	&	  1		&	3.04689   & Focus   \\
	7	&	0.818957	&	0.818957	&	  1		&	7.36865   & Center	  \\	
        \br
\end{tabular}
\end{table}

\begin{figure}[htb]
    \centering
  \includegraphics[width=85 mm,height= 70 mm]{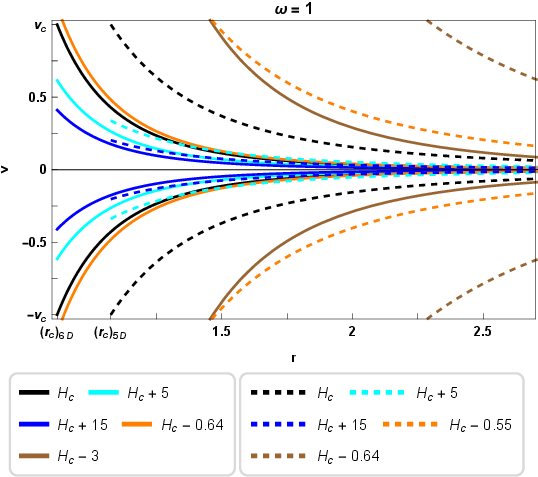} 
\caption{Contour plots of  Hamiltonian $\mathcal{H}$ with 5D (Dashed Curve) and 6D (Line curve) BHs for the parameters $\omega = \frac{1}{2}$, $M = 1.5$ and $q = 1$. The black curve corresponds to the critical Hamiltonian $\mathcal{H}_c$. The Blue curve corresponds to $\mathcal{H} > \mathcal{H}_c $, and the Orange curve represents $\mathcal{H} < \mathcal{H}_c$  } 
 	\label{fig:usf1}
\end{figure}

In the case of the ultra-stiff fluid (USF), the behaviour of the dynamical system can be studied using the equations (\ref{HS1}) and (\ref{HS2}), which follow for all the BHs. Since the Hamiltonian $\mathcal{H}$ given by equation (\ref{H1}) is constant of motion, we observe that $v$ follows a law given by  $ v \propto \frac{1}{r^{(D-2)}}$. In figure (\ref{fig:usf}), we plot the phase space diagram of the dynamical system for BHs that are found in the even or odd dimensional space-time; we displayed the result with different parameters in table \ref{tab:usf}. The upper region of the curves, where $v > 0$, represents the fluid outflow or particle emission and the lower region of the curves, where $v<0$, represents the fluid accretion.  We plot the contour corresponding to the Hamiltonian given in (\ref{H1}) in figure (\ref{fig:usf}) which leads to two types of physical flows ($|v| <1$): (i) critical subsonic accretion ($v<0$) or critical subsonic flow out ($v>0$) (along black curves) and (ii) purely subsonic accretion ($v<0$) or purely subsonic flow out ($v>0$) (along cyan and blue curves). Fluids flowing along the orange and brown curves are unphysical since the speed of the flow exceeds the speed of light on some proportions of the curves. We also note from the figure (\ref{fig:usf1}) that the accretion of USF is independent of dimensions, but the critical values are found to shift towards the event horizon, and the critical Hamiltonian ($\mathcal{H}_c$) increases as space-time dimensions are increased, which is tabulated in table \ref{tab:usf}. It is also noted that the radial velocity of accretion decreases as the distance of the fluid increases as the horizon increases and at $r \to \infty$, $v_{\infty} = 0$.

\begin{figure*}[h]
\centering
\begin{tabular}{cc}
  \includegraphics[width=0.45\textwidth,height= 0.35\textwidth]{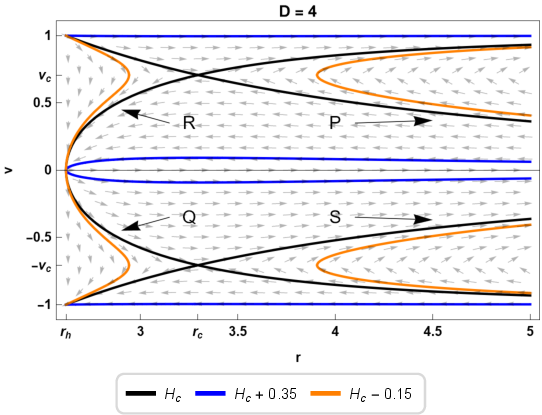} &   \includegraphics[width=0.45\textwidth,height= 0.35\textwidth]{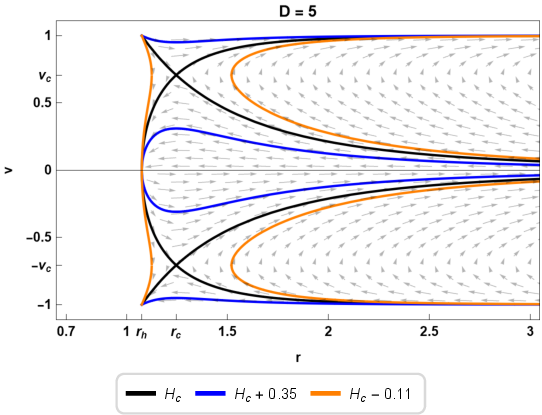}\\[6pt]
  \includegraphics[width=0.45\textwidth,height= 0.35\textwidth]{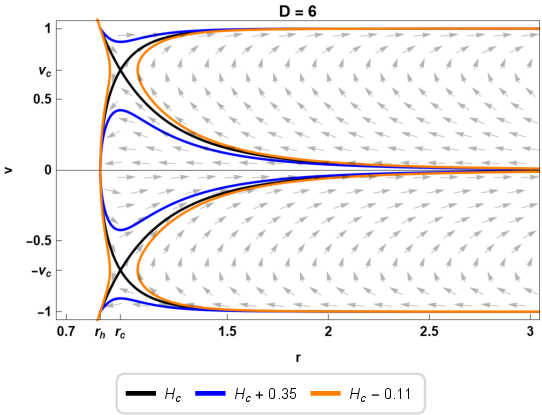} &   \includegraphics[width=0.45\textwidth,height= 0.35\textwidth]{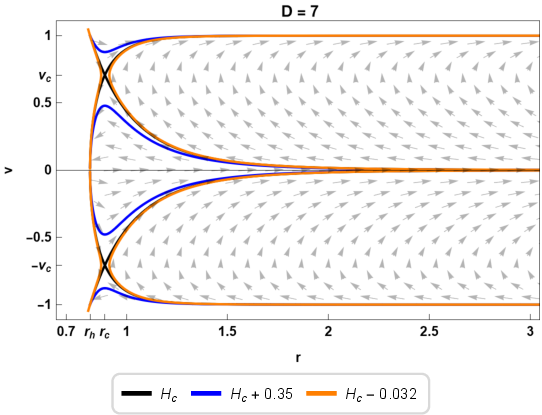}\\[6pt]
\end{tabular}
\caption{Contour plots of  Hamiltonian $\mathcal{H}$   with different dimensional BHs for the parameters $\omega = \frac{1}{2}$, $M = 1.5$ and $q = 1$. The black curve corresponds to the critical Hamiltonian $\mathcal{H}_c$. The Blue curve corresponds to $\mathcal{H} > \mathcal{H}_c $, and the Orange curve represents $\mathcal{H} < \mathcal{H}_c$.  } 
 	\label{fig:urf}
\end{figure*}

\subsection{Case II : Ultra Relativistic Fluid ($\omega = \frac{1}{2}$)}

The ultra-relativistic fluid (URF) is classified by the EoS $p = \frac{\rho}{2}$, where the state parameter is $\omega = \frac{1}{2}$. Comparing equations (\ref{uc1}) and (\ref{uc2}), the critical radius $r_c$ can be obtained, which yields from the equation $(1-\omega) \,r_c \,f(r_c)_{,r_c} - 2(D-2)\omega f(r_c) = 0 $. We obtain the following
\begin{equation}
\centering
	\label{criticalradius}
	\frac{32\pi^2q^2 r^{6-D} (D-3 + \omega )}{(D-3)\,\Omega_{D-2} } +  \omega\, r^D(D-2)^2\,  \Omega_{D-2}  - 8M \pi \,r^3\, (D-3+\omega(D-1)) = 0.
\end{equation}
Using the critical radius  $r_c$ in equation (\ref{vsquared}), we determine the critical value of the radial velocity ($v_c$) of the fluid flow. Thereafter, making use of the critical values ($r_c,\pm v_c$), we determine the critical Hamiltonian for ultra-relativistic fluids. Thus, the general Hamiltonian for URF becomes
\begin{equation}
	\label{H2}
	\mathcal{H} = \frac{f(r)^{\frac{1}{2}}}{r^{(D-2)}\,|v|\,(1-v^2)^{\frac{1}{2}}}.
\end{equation}
It is evident from equation (\ref{H2}) that the point ($r,v^2$) = ($r_h,1$) is not a Critical Point of the dynamical system. Now, for a given value of the Hamiltonian $\mathcal{H} = \mathcal{H}_c$, the solution can be obtained for $v$ using equation (\ref{H2}) as
\begin{equation}
	v^2 = \frac{1 \pm \Big( 1-4 \,x(r) \Big)^{\frac{1}{2}}}{2},
\end{equation} 
where, $x(r) = \frac{f(r)}{  \mathcal{H}^2 r^{2(D-2)})}$. Using the above equation, numerically, we analyse the phase space diagram of the dynamical system given in equations (\ref{HS1}) and (\ref{HS2}) for a given value of the Hamiltonian.

 \begin{figure}[ht]
    \centering
  \includegraphics[width=90 mm,height= 70 mm]{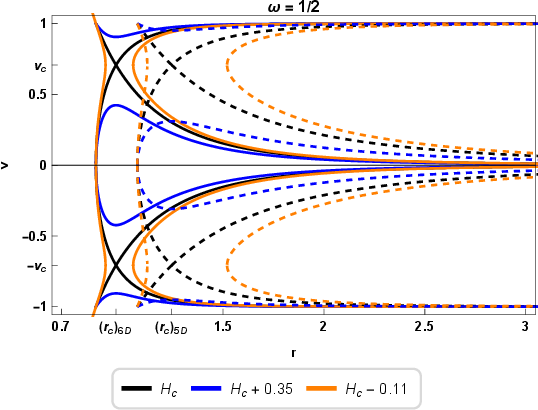} 
\caption{Contour plots of  Hamiltonian $\mathcal{H}$ with 5D (Dashed Curve) and 6D (Line curve) BHs for the parameters $\omega = \frac{1}{2}$, $M = 1.5$ and $q = 1$. The black curve corresponds to the critical Hamiltonian $\mathcal{H}_c$. The Blue curve corresponds to $\mathcal{H} > \mathcal{H}_c $, and the Orange curve represents $\mathcal{H} < \mathcal{H}_c$.  } 
 	\label{fig:urf1}
\end{figure}
\begin{table}[t] 
\centering
\caption{\label{tab:urf}}
Values of $r_c$, $v_c$, and $H_c$ at the sonic point for ultra-relativistic fluid ($\omega  = \frac{1}{2}$) for even and odd dimensional BHs with suitable parameters $M = 1.5$ and $q = 1$.
\begin{tabular}{c c c c c c}
        \br
	D	&	$r_h$	    &	$r_c$      	&	$v_c$    	&	$H_c$    & Equilibrium point type	  \\
	\mr
	4	&	2.618030	&	3.29473	    &	 0.707105	&	0.0785091 & Saddle	\\
	5	&	1.075370	&	1.24605	    &	 0.707105	&	0.502195  & Saddle	\\
	6	&	0.869997	&	0.969709	&	 0.707106	&	1.15391  & Saddle	\\
	7	&	0.818957	&	0.892416	&	 0.707107	&	1.84972  & Saddle	\\	
        \br
\end{tabular}
\end{table}

In figure (\ref{fig:urf}), the phase space diagrams of the dynamical system are plotted for the critical  Hamiltonian $\mathcal{H}_c$. In the dynamical system, we also obtained the contours for the following: (i)  $\mathcal{H} > \mathcal{H}_c$ and (ii) $\mathcal{H} < \mathcal{H}_c$ for even and odd dimensional BHs. Numerical values for critical radius, critical velocity and critical Hamiltonian are obtained and tabulated in table   \ref{tab:urf} with BH parameters $M = 1.5$ and $q = 1$. We analyze the trajectories of the accretion fluids by a contour plot and found  four different types of physical fluid motion, which can be classified as :

 \begin{itemize}
 
     \item Type-I: When $\frac{v}{r}$ is continuous at the Critical Points, we get a critical supersonic accretion up to ($r_c,-v_c$) followed by a critical subsonic accretion at ($r_h,0$), then one obtains a critical subsonic outflow upto ($r_c,v_c$) followed by a critical supersonic outflow of motion (Correspond to the curves Q and R, respectively) after that;

     \item {\bf Type-II: Assuming continuity at the Critical Points, we get a critical subsonic accretion until ($r_c,-v_c$) followed by a critical supersonic accretion, which ends up at the horizon and a critical supersonic outflow that originates from the horizon up to ($r_c,v_c$) followed by a critical subsonic outflow of fluid (Correspond to the curves S and P, respectively);}

     \item Type-III: Purely supersonic accretion corresponds to  $-1 < v < -v_c $ (bottom blue curve) and  for purely supersonic outflow  for $v_c < v < 1$ (Top blue curve);

     \item Type-IV:  Purely subsonic accretion for $-v_c< v < 0$ and purely subsonic outflow for $0 < v < v_c$ (blue curves in the middle). In this case, the fluid reaches the horizon with a vanishing speed, keeping a constant value of the Hamiltonian.
 \end{itemize}
 
For $H_c$ we get four different curves which are marked by $P$, $Q$, $R$, and $S$ in the top left panel of the figure (\ref{fig:urf}), for $D=4$. Similar features can be observed for other dimensions $D=5,\; 6, \;7$.  Considering the fact that the accretion does not modify the intrinsic properties of a BH, we have analyzed accretion processes here. The accretion flow is non-geodesic in nature. If we draw a vertical line through the critical point $r=r_c$, it is found that there are two branches of the orange contour on either side. For the branches on the right of the vertical line $r=r_c$, the fluid flow along the segment of that branch is neither an accretion nor an outflow and is unphysical. In the leftmost branch of the orange curve, the accretion begins from the leftmost point of the branch till the horizon, where the speed of the fluid vanishes; this is followed by an outflow of fluid back to the same point. Such a flow can be realised if we consider the sink and source at the same point, which is again unphysical. We plot figure (\ref{fig:urf1}) to compare the accretion behaviour at two different dimensions, namely, $D=5$ and $D=6$, and found that the critical points are shifted towards the event horizon as the spacetime dimensions increase. The contours for accretion onto a BH are found to move near the event horizon as the spacetime dimension increases.


\subsection{Case III : Radiation Fluid ($\omega = \frac{1}{3}$)}
 We consider Radiation fluid (RF) in this section for which the EoS reduces to $p = \frac{\rho}{3}$, and the state parameter is $\omega = \frac{1}{3}$. In equation (\ref{criticalradius}), we use $\omega$ and obtain the critical radius $r_c$ from the equation :
\begin{equation}
	\label{cr3}
	\frac{32\pi^2\,q^2 \,r^{6-D} ( 3D - 8 )}{(D-3)\,\Omega_{D-2} } +   r^D\,(D-2)^2  \Omega_{D-2} - 16M \pi\, r^3 (2D-5) = 0.
\end{equation}
\begin{figure*}[t]
\centering
\begin{tabular}{cc}
  \includegraphics[width=0.45\textwidth,height= 0.35\textwidth]{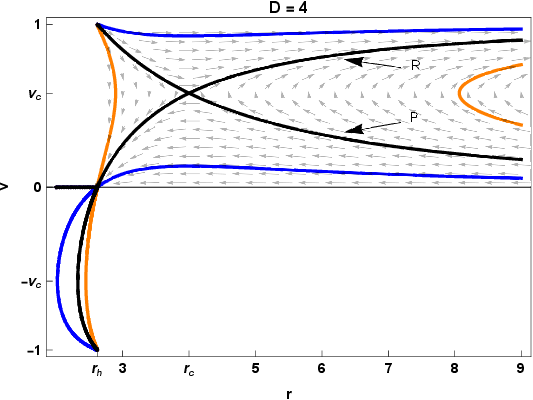} &   \includegraphics[width=0.45\textwidth,height= 0.35\textwidth]{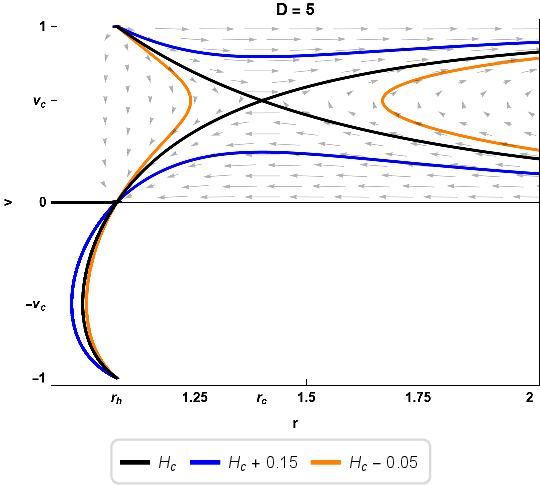}\\[6pt]
  \includegraphics[width=0.45\textwidth,height= 0.35\textwidth]{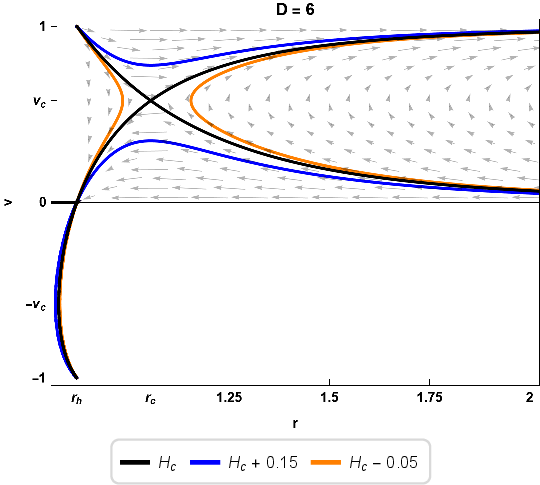} &   \includegraphics[width=0.45\textwidth,height= 0.35\textwidth]{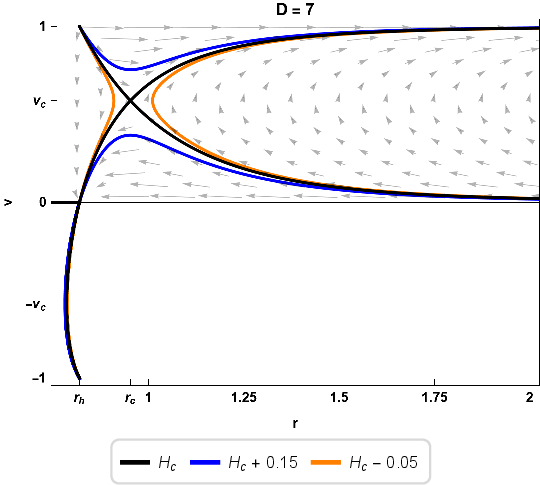}\\[6pt]
\end{tabular}
\caption{Contour plots of  Hamiltonian $\mathcal{H}$   with different dimensional BHs for the parameters $\omega = \frac{1}{3}$, $M = 1.5$ and $q = 1$. The Black curve corresponds to the critical Hamiltonian $\mathcal{H}_c$. The Blue curve corresponds to $\mathcal{H} > \mathcal{H}_c $, and the Orange curve represents $\mathcal{H} < \mathcal{H}_c$.  } 
 	\label{fig:rf}
\end{figure*} 
                                    
\begin{table}[ht] 
\centering
\caption{\label{tab:rf}}
Values of $r_c$, $v_c$, and $H_c$ at the sonic point for radiation fluid ($\omega  = \frac{1}{3}$) for even and odd dimensional BHs with suitable parameters $M = 1.5$ and $q = 1$.
\begin{tabular}{c c c c c c}
        \br
	D	&	$r_h$	    &	$r_c$	&	$v_c$	    &	$H_c$     & Equilibrium point type	  \\
	\mr
	4	&	2.618030	&	4		&	 0.57735	&	0.137063  & Saddle	\\
	5	&	1.075370	&	1.40048	&	 0.577351	&	0.510782  & Saddle	\\
	6	&	0.869997	&	1.05442	&	 0.577355	&	0.915431  & Saddle	\\
	7	&	0.818957	&	0.952629&	 0.57735	&	1.27292   & Saddle	\\	
        \br
\end{tabular}
\end{table}

 \begin{figure}[h]
    \centering
  \includegraphics[width=90 mm,height= 75 mm]{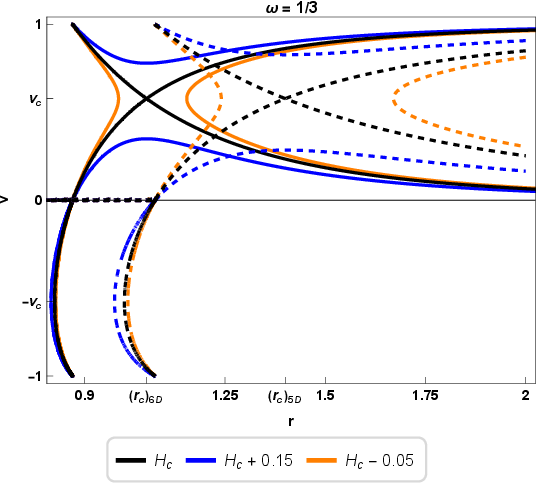} 
\caption{Contour plots of  Hamiltonian $\mathcal{H}$ with 5D (Dashed Curve) and 6D (Line curve) BHs for the parameters $\omega = \frac{1}{3}$, $M = 1.5$ and $q = 1$. The black curve corresponds to the critical Hamiltonian $\mathcal{H}_c$. The Blue curve corresponds to $\mathcal{H} > \mathcal{H}_c $, and the Orange curve represents $\mathcal{H} < \mathcal{H}_c$.  } 
 	\label{fig:rf1}
\end{figure}

The critical radial velocity ($v_c$) of the fluid is obtained from equation (\ref{vsquared}), and the general Hamiltonian $\mathcal{H}$ given in equation (\ref{finalhamiltonian}) becomes
\begin{equation}
	\label{Hrf}
	\mathcal{H} = \frac{f(r)^{\frac{2}{3}}}{r^{\frac{2(D-2)}{3}}\,|v|^{\frac{2}{3}}\,(1-v^2)^{\frac{2}{3}}}.
\end{equation}
 The contour plots of the Hamiltonian $\mathcal{H}$ for even and odd dimensional BHs are shown in figure (\ref{fig:rf}) for critical values of the Hamiltonian ($\mathcal{H}_c$), and two other cases, {\it i.e.}, $\mathcal{H} > \mathcal{H}_c$ and $\mathcal{H} < \mathcal{H}_c$. The parameters related to the plot are tabulated in table  \ref{tab:rf}. The radiation fluid shows three different types of physical fluid motion, which can be classified as :
\begin{itemize}
    \item Type-I: Assuming the continuity of $\frac{v}{r}$ at the Critical point $(r_c,v_c)$, we get a supersonic outflow until $(r_c,v_c)$, followed by a subsonic outflow of fluid (corresponds to the curve P);
    \item Type-II: Assuming the continuity at the Critical point $(r_c,v_c)$, we get a subsonic outflow until $(r_c,v_c)$, followed by a supersonic outflow (corresponds to the curve R);
    \item Type-III: Purely supersonic outflow of radiative fluid in the region  $v_c < v < 1$, and purely subsonic outflow for $0 < v < v_c$.
\end{itemize}
In figure (\ref{fig:rf}), Type-I and Type-II correspond to the critical Hamiltonian (black curve), whereas, Type-III correspond to $\mathcal{H} > \mathcal{H}_c$ (blue curves). The orange curves corresponding $\mathcal{H} < \mathcal{H}_c$ are unphysical. It is evident that radiative fluid exhibits similar flow behaviour for $|v| >1$ and unphysical behaviour for $|v| < 1$. From figure (\ref{fig:rf1}), we further note that as the dimension increases, the critical radius ($r_c$) decreases, but the critical Hamiltonian ($\mathcal{H}_c$) increases but without any new result.


\begin{figure*}[ht]
\centering
\begin{tabular}{cc}
  \includegraphics[width=0.45\textwidth,height= 0.35\textwidth]{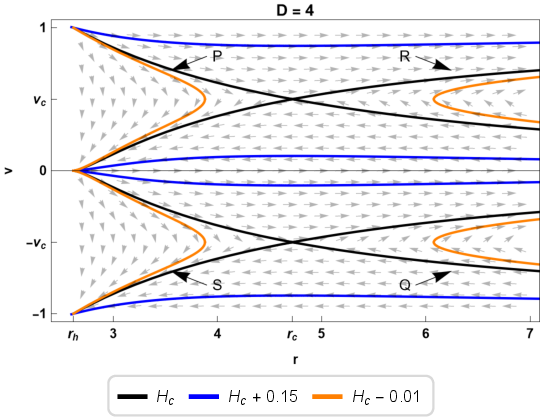} &   \includegraphics[width=0.45\textwidth,height= 0.35\textwidth]{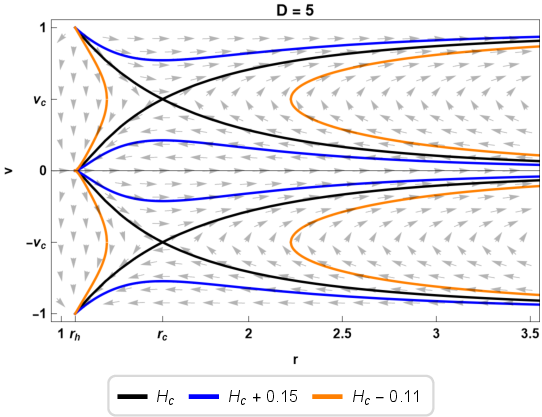}\\[6pt]
  \includegraphics[width=0.45\textwidth,height= 0.35\textwidth]{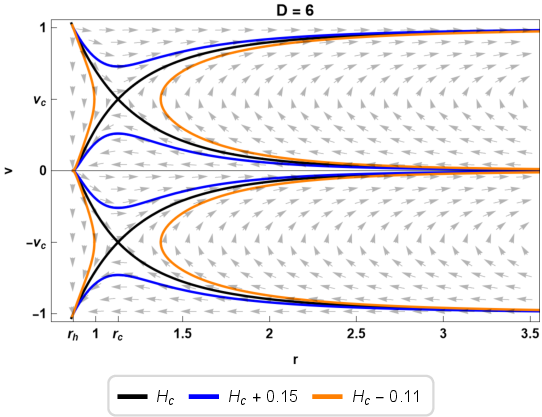} &   \includegraphics[width=0.45\textwidth,height= 0.35\textwidth]{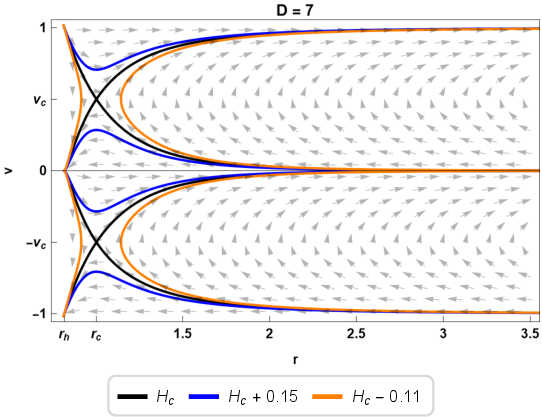}\\[6pt]
\end{tabular} 
\caption{Contour plots of  Hamiltonian $\mathcal{H}$   with different dimensional BHs for the parameters $\omega = \frac{1}{4}$, $M = 1.5$ and $q = 1$. The black curve corresponds to the critical Hamiltonian $\mathcal{H}_c$. The Blue curve corresponds to $\mathcal{H} > \mathcal{H}_c $, and the Orange curve represents $\mathcal{H} < \mathcal{H}_c$.  } 
 	\label{fig:srf}
\end{figure*}
                                    
\subsection{Case IV : Sub-Relativistic Fluid ($\omega = \frac{1}{4}$)}

We consider here sub-relativistic fluid (SRF), the EoS becomes $p = \frac{\rho}{4}$, where the state parameter is  $\omega = \frac{1}{4}$. Using the value of $\omega$ in equation (\ref{criticalradius}), we obtain the critical radius $r_c$ from the following equation
\begin{equation}
	\label{cr4}
	\frac{32\pi^2\,q^2 r^{6-D} ( 4D - 11 )}{(D-3)\Omega_{D-2} } +   r^D\,(D-2)^2  \,\Omega_{D-2} - 8M \pi\, r^3 (5D-13) = 0.
\end{equation}
Using the critical radius $r_c$, we determine $v_c$ from equation (\ref{vsquared}), and the general Hamiltonian $\mathcal{H}$ from equation (\ref{finalhamiltonian}) can be expressed as
\begin{equation}
	\label{H4}
	\mathcal{H} = \frac{f(r)^{\frac{3}{4}}}{r^{\frac{(D-2)}{2}}\,|v|^{\frac{1}{2}}\,(1-v^2)^{\frac{3}{4}}}.
\end{equation}

 \begin{table}[h] 
 \centering
\caption{\label{tab:srf}}
Values of $r_c$, $v_c$, and $H_c$ at the sonic point for sub-relativistic fluid ($\omega  = \frac{1}{4}$) for even and odd dimensional BHs with suitable parameters $M = 1.5$ and $q = 1$.
\begin{tabular}{c c c c c c}
        \br
	D	&	$r_h$	    &	$r_c$	&	$v_c$	&	$H_c$     & Equilibrium point type \\
	\mr
	4	&	2.618030	&	4.72038	&	 0.5	&	0.190241  & Saddle	\\
	5	&	1.075370	&	1.54139	&	0.50002	&	0.535409  & Saddle		\\
	6	&	0.869997	&	1.12818	&	0.499997&	0.843824  & Saddle		\\
	7	&	0.818957	&	1.00371	&	0.50001	&	1.09026   & Saddle		\\	
        \br
\end{tabular}
\end{table}

\begin{figure}[ht]
    \centering
  \includegraphics[width=90 mm,height= 70 mm]{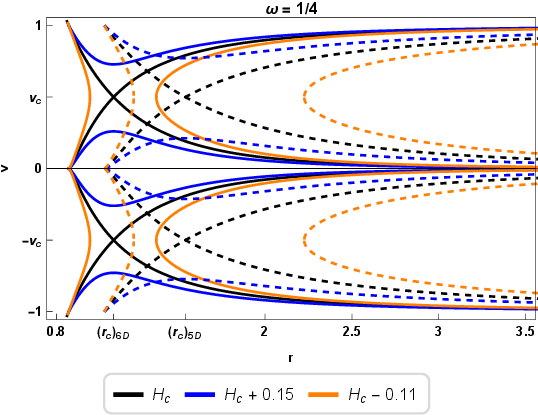} 
\caption{Contour plots of  Hamiltonian $\mathcal{H}$ with 5D (Dashed Curve) and 6D (Line curve) BHs for the parameters $\omega = \frac{1}{4}$, $M = 1.5$ and $q = 1$. The black curve corresponds to the critical Hamiltonian $\mathcal{H}_c$. The Blue curve corresponds to $\mathcal{H} > \mathcal{H}_c $, and the Orange curve represents $\mathcal{H} < \mathcal{H}_c$.  } 
 	\label{fig:srf1}
\end{figure}

The contours of the Hamiltonian $\mathcal{H}$ are plotted in figure (\ref{fig:srf}) for both the even and odd dimensional BHs. The contours are also plotted for the critical values of the Hamiltonian ($\mathcal{H}_c$) along with $\mathcal{H} > \mathcal{H}_c$ and $\mathcal{H} < \mathcal{H}_c$. The physical behaviour of the accretion flow is found similar in higher-dimensional BHs. We found that the critical values of the parameters may change, which are tabulated in table \ref{tab:srf}. It is evident from the trajectories of the sub-relativistic fluid plotted in figure (\ref{fig:srf}) that four types of physical fluid motion are possible for the BH which are drawn by the black and blue curves, similar to that obtained in the ultra-relativistic fluid shown in figure (\ref{fig:urf}). We observe both purely supersonic and subsonic accretion for $-1<v<v_c$ and $0>v>v_c$, respectively (Blue Curves).  We also found purely supersonic and subsonic fluid out-flow for $1>v>v_c$ and $0<v<v_c$, respectively (Blue Curves). Assuming the continuity at the Critical points, it is found that critical subsonic accretion is followed by a critical supersonic accretion up to the horizon, where it vanishes (corresponds to S curve), and a critical supersonic outflow from the horizon followed by a critical subsonic outflow (corresponds to P curve). Critical supersonic accretion followed by critical subsonic accretion up to ($r_h,0$), then a critical subsonic outflow followed by a critical supersonic outflow is also visible (corresponds to Q and R curves). The curves marked by $P$, $Q$, $R$, and $S$ for $D=4$ apply to the other dimensions we get similar behaviour. We notice unphysical fluid motion corresponding to the fluid motion shown by the orange curves in all dimensions. It is found from figure (\ref{fig:srf1}) that the critical radius ($r_c$) decreases, and the critical Hamiltonian ($\mathcal{H}_c$) increases as dimensions are increased. \\

\begin{figure*}[ht]
\centering
\begin{tabular}{cc}
  \includegraphics[width=0.45\textwidth,height= 0.33\textwidth]{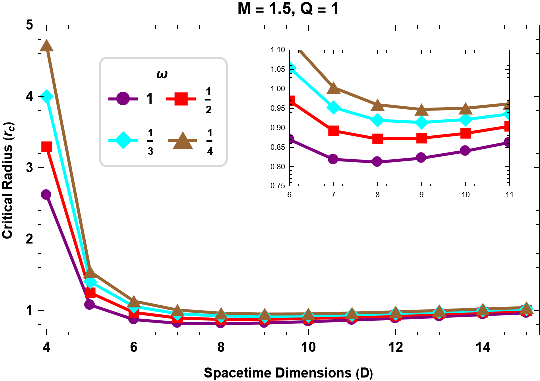} &   \includegraphics[width=0.45\textwidth,height= 0.33\textwidth]{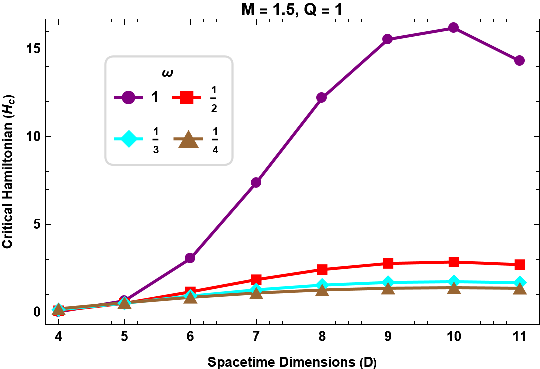}\\[6pt]
  \includegraphics[width=0.45\textwidth,height= 0.33\textwidth]{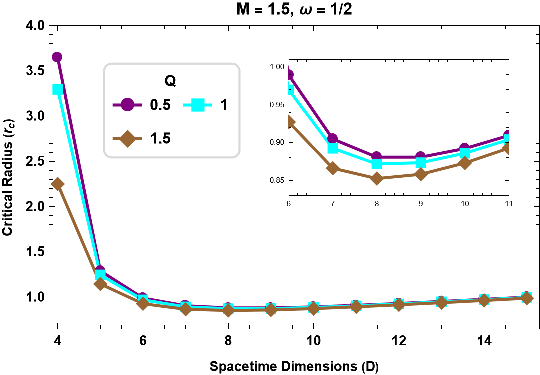} &   \includegraphics[width=0.45\textwidth,height= 0.33\textwidth]{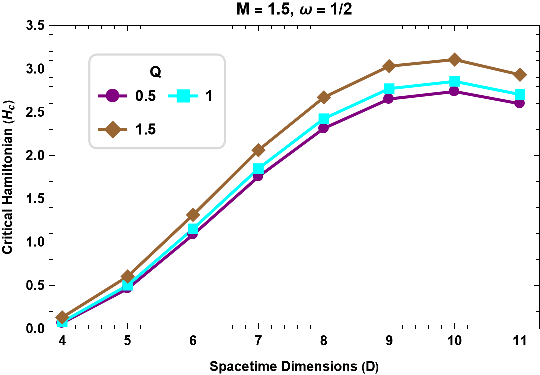}\\[6pt]
\end{tabular} 
\caption{Comparison of the critical radius ($r_c$) and the critical Hamiltonian ($H_c$) against spacetime dimension ($D$) for different values of the state parameter $\omega$ with BH parameters  $M = 1.5$, $q = 1$ (top). Comparison of the critical radius ($r_c$) and the critical Hamiltonian ($H_c$) against spacetime dimension ($D$) for different values of the BH charge parameter ($q$) with state parameter $\omega = \frac{1}{2}$, and BH mass parameter $M = 1.5$ (bottom) } 
 	\label{fig:drh}
\end{figure*}

 The change pattern of the critical radius ($r_c$) and the critical Hamiltonian ($H_c$) against the spacetime dimensions is plotted in figure (\ref{fig:drh}) for isotropic fluid with different state parameters. We note that the critical radius is decreasing as the spacetime dimension increases. However, the critical Hamiltonian, which is the total energy of the autonomous dynamical system, increases with an increase in the spacetime dimension for all values of the state parameters. In table \ref{tab:iso}, we have shown the flow behaviour for different fluids through charts.


 \begin{table}[ht] 
 \centering
\caption{\label{tab:iso}}
Flow behaviour for isotropic fluid with different values of $\omega$ with BH parameters $M = 1.5$ and $q = 1$.
\begin{tabular}{c | c | c}
        \br
	$\omega$  & Types	&	Flow behaviour  \\
	\mr
	           &   I	 &	$\mathcal{H}$ = $\mathcal{H}_c$	: Critical subsonic                          accretion for $v <0$ and outflow for $v>0$		\\
		1         &   II	&	$\mathcal{H}$ $>$ $\mathcal{H}_c$	: Purely subsonic                            accretion for $v <0$ and outflow for $v>0$		\\
		         &   III   &	$\mathcal{H}$ $<$ $\mathcal{H}_c$	: Unphysical flow\\
	\mr
	           &   I	 &	$\mathcal{H}$ = $\mathcal{H}_c$	: Critical supersonic          accretion until ($r_c,-v_c$), followed by critical subsonic flow \\
                  &         &   from ($r_c,-v_c$) to ($r_c,v_c$), then critical supersonic outflow after ($r_c,v_c$)	\\
	           &   II	 &	$\mathcal{H}$ = $\mathcal{H}_c$	: Critical subsonic                  accretion until ($r_c,-v_c$), then critical supersonic; \\
            $\frac{1}{2}, \frac{1}{4}$  &         &    Critical supersonic outflow                      until ($r_c,-v_c$), then critical subsonic	\\
		         &   III	&	$\mathcal{H}$ $>$ $\mathcal{H}_c$	: Purely supersonic                        accretion for $-1< v <v_c$ and outflow for $v_c < v < 1$\\
		         &   IV   &	$\mathcal{H}$ $>$ $\mathcal{H}_c$	: Purely subsonic                      accretion for $-v_c < v < 0$ and outflow for $0< v <v_c$\\
		         &   V   &	$\mathcal{H}$ $<$ $\mathcal{H}_c$	: Unphysical flow\\
        \mr
	           &   I	 &	$\mathcal{H}$ = $\mathcal{H}_c$	: Critical supersonic                  outflow until ($r_c,v_c$), then critical subsonic\\
	           &   II	 &	$\mathcal{H}$ = $\mathcal{H}_c$	: Critical subsonic                  outflow until ($r_c,v_c$), then critical subsonic \\
    $\frac{1}{3}$ &   III	&	$\mathcal{H}$ $>$ $\mathcal{H}_c$	: Purely supersonic                        outflow for $v_c < v < 1$; \\
                  &         &    Purely subsonic outflow for $0 < v < v_c$\\
		         &  IV     &	$\mathcal{H}$ $<$ $\mathcal{H}_c$	: Unphysical flow\\
        \br
\end{tabular}
\end{table}
                                    

\section{Applications to Polytropic fluids accretion }
\label{sec:5}
We consider Chaplygin gas to be the prototype of polytropic fluid. Chaplygin gas, which is extended to modified Chaplygin gas, is found to have an important role in astrophysics \cite{wu07,bcp16,ds16}. The EoS of the modified Chaplygin gas is
\begin{equation}
	\label{chaplygin}
	p = An - \frac{B}{\varrho^{\alpha}},
\end{equation}
where $A$ and $B$ are constants and ($0<\alpha <1$). If we substitute $A=0$, $B=-\kappa$, and $\alpha = - \Gamma$, we get the polytropic EoS, {\it i.e.}
\begin{equation}
	\label{polyeos}
	p = \kappa\, \varrho^{\Gamma},
\end{equation}
where $\kappa$ and $\Gamma$ are arbitrary constants and $\Gamma > 0$ for ordinary matter. Using the first law of thermodynamics in equation (\ref{FLT}) and equation  (\ref{polyeos}), we can obtain the following
\begin{equation}
	\label{polyenthalpy1}
	h = m + \frac{\Gamma}{\Gamma-1} \kappa \,\varrho^{\Gamma-1},
\end{equation}
where $m$ is the integration constant. Using equation (\ref{polyenthalpy1}) in equation (\ref{sos}), yields
\begin{equation}
	\label{polysos}
	a^2 = \frac{(\Gamma-1)\,\mathcal{X}}{\mathcal{X} + m\, (\Gamma-1)},\;\;\;\; (\mathcal{X} \equiv \kappa\, \Gamma \,\varrho^{\Gamma-1}).
\end{equation}
                                   
Now, using equation (\ref{numberdensityratio}) in equation (\ref{polyenthalpy1}), we arrive at
\begin{equation}
	\label{polyenthalpy2}
	h = m \Bigg[ 1 + \mathcal{Y}\, \Bigg( \frac{1-v^2}{r^{2(D-2)}\,v^2\, f(r)} \Bigg)^{\frac{\Gamma-1}{2}} \Bigg],
\end{equation}
where,
\begin{equation}
	\label{Y}
	\mathcal{Y} = \frac{\kappa \,\Gamma \,\varrho_c^{\Gamma-1}}{m\,(\Gamma-1)} \Bigg(\frac{r_c^{2D-3}\,f(r_c)_{,r_c}}{2(D-2)} \Bigg)^{\frac{\Gamma-1}{2}} \ =\ const..
\end{equation}

\begin{figure*}[t]
\centering
\begin{tabular}{cc}
  \includegraphics[width=0.45\textwidth,height= 0.35\textwidth]{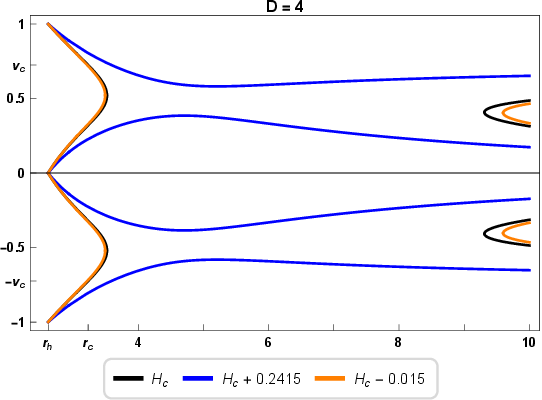} &   \includegraphics[width=0.45\textwidth,height= 0.35\textwidth]{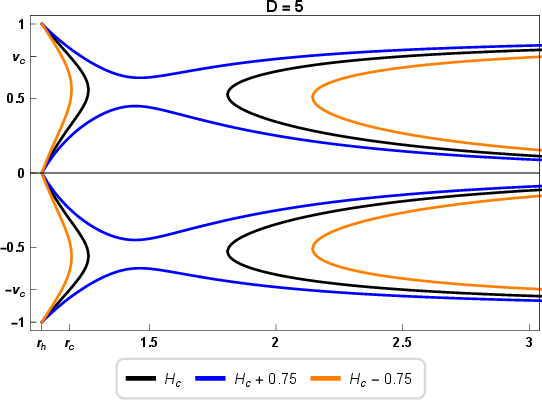}\\[6pt]
  \includegraphics[width=0.45\textwidth,height= 0.35\textwidth]{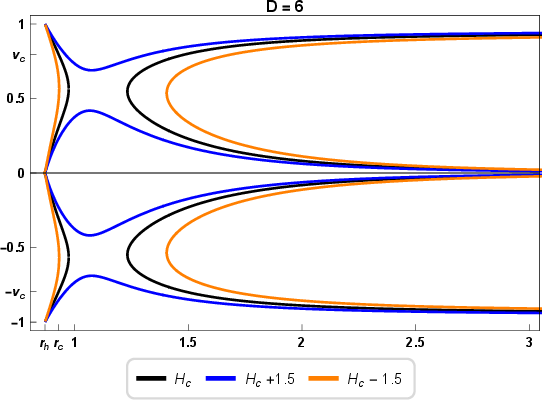} &   \includegraphics[width=0.45\textwidth,height= 0.35\textwidth]{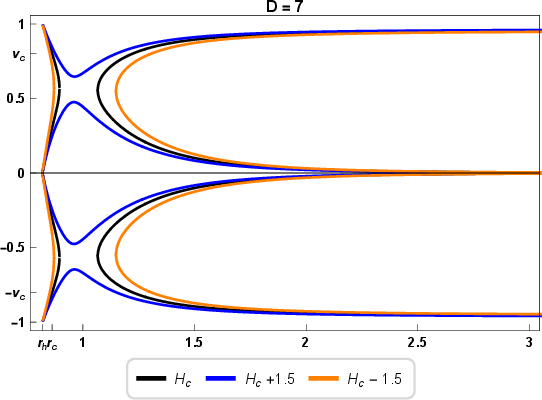}\\[6pt]
\end{tabular} 
\caption{Contour plots of  Hamiltonian $\mathcal{H}$  with different dimensional BHs for the parameters  $M = 1.5$, $q = 1$, $\Gamma = 5/3$ and $\mathcal{Y} = 10$. The black curve corresponds to the critical Hamiltonian $\mathcal{H}_c$. The Blue curve corresponds to $\mathcal{H} > \mathcal{H}_c $, and the Orange curve represents $\mathcal{H} < \mathcal{H}_c$.  } 
 	\label{fig:poly1}
\end{figure*}

Substituting equation (\ref{polyenthalpy2}) in equation (\ref{hamiltonian}), the Hamiltonian for the polytropic EoS can be obtained as
\begin{equation}
	\label{polyHamiltonian}
	\mathcal{H} = \frac{f(r)}{1-v^2} \Bigg[ 1 + \mathcal{Y}\, \Bigg( \frac{1-v^2}{r^{2(D-2)}\,v^2 f(r)} \Bigg)^{\frac{\Gamma-1}{2}} \Bigg]^2,
\end{equation}
where the integration constant $m^2$ is absorbed into the re-definition of $(\bar{t},\mathcal{H})$. We note that the square parenthesis in equation (\ref{polyHamiltonian}) is positive; while the coefficient $\frac{f(r)}{1-v^2}$ diverges as $r \to \infty$ ($0\leq 1-v^2 <1$). 
Now, using equation (\ref{numberdensityratio}), we rewrite the expression for $\mathcal{X}$ as
\begin{equation}
    \mathcal{X} = m\,(\Gamma - 1)\,\mathcal{Y}\, \Bigg( \frac{1-v^2}{r^{2(D-2)}\,v^2 f(r)} \Bigg)^{\frac{\Gamma-1}{2}}.
\end{equation}
Using the above equation, we arrive at
\begin{equation}
    \label{polysoundspeed}
    a^2 = \mathcal{Y}\, (\Gamma - 1 - a^2) \Bigg( \frac{1-v^2}{r^{2(D-2)}\,v^2 f(r)} \Bigg)^{\frac{\Gamma-1}{2}},
\end{equation}
which along with equation (\ref{sosCP}) take the following expressions at critical points
\begin{equation}
    v_c^2 = \mathcal{Y}\, (\Gamma - 1 - v_c^2) \Bigg( \frac{1-v^2}{r^{2(D-2)}\,v^2 f(r)} \Bigg)^{\frac{\Gamma-1}{2}}, \nonumber
\end{equation} 
\begin{equation}
    v_c^2 = \frac{r_c\,f(r_c)_{,r_c}}{r_c\,f(r_c)_{,r_c} + 2(D-2)f(r_c)}.
\end{equation}

 \begin{table}[t] 
    \centering
\caption{\label{tab:poly}}
Values of $r_c$, $v_c$, and $H_c$ at the sonic point for polytropic fluid for even and odd dimensional BHs with suitable parameters $M = 1.5$, $q = 1$, $\Gamma = 5/3$ and $\mathcal{Y} = 10$.\\
\begin{tabular}{c c c c c}
        \br
	D	&	$r_h$	&	$r_c$	&	$v_c$	&	$H_c$	  	\\
	\mr
	4	&	2.618030	&	3.23074	&	 0.72402	&	1.64473		\\
	5	&	1.075370	&	1.18556	&	0.782318  	&	5.11019		\\
	6	&	0.869997	&	0.930106 &	0.7951   	&	8.86145		\\
	7	&	0.818957	&	0.862172 &	0.800165	&	12.1865		\\	
        \br
\end{tabular}
\end{table}
 
In figure (\ref{fig:poly1}) we plot the contour of the Hamiltonian for the BH parameters $M = 1.5$, $q = 1$, $\Gamma = 5/3$ and $\mathcal{Y} = 10$ for different dimensions. The contour is plotted for the critical values of the Hamiltonian ($H_c$) (black curve) along with the Hamiltonian $\mathcal{H} > \mathcal{H}_c$ (blue curve) and $\mathcal{H} < \mathcal{H}_c$ (orange curve). A critical and relativistic accretion from the left-most point of the black curve is observed until the horizon, where the speed vanishes, followed by a critical flow-out. However, this kind of flow is unphysical, since it represents a source and sink at the same point. We also found two other fluid flows, one of which is relativistic and supersonic near the horizon, which afterwards becomes subsonic in the midway and then becomes supersonic again at spatial infinity (uppermost and lowermost branches of the blue curve) and the other fluid flow is purely subsonic accretion and flow-out with vanishing speed at the horizon and non-zero at the spatial infinity (blue curves within $-v_c < v < v_c$). The rightmost black and orange curves are unphysical. It is also noted from table \ref{tab:poly} that the critical radius decreases and the critical Hamiltonian is found to increase with an increase in the spacetime dimension. In table \ref{tab:polyall}, we have shown the flow behaviour for polytropic fluid through charts.

 \begin{table}[ht] 
 \centering
\caption{\label{tab:polyall}}
Flow behaviour for polytropic fluid ($\gamma = 5/3$) with BH parameters $M = 1.5$ and $q = 1$.\\
\begin{tabular}{ c | c}
        \br
	 Types	&	Flow behaviour  \\
	\mr
	I	 &	$\mathcal{H}$ = $\mathcal{H}_c$	: Unphysical Flow	\\
		II	 &	$\mathcal{H}$ $>$ $\mathcal{H}_c$: Non-critical supersonic accretion/outflow near horizon, becomes subsonic midway  \\
             &  and then supersonic accretion/outflow at spatial infinity (top and bottom blue curve) \\
		III	 &	$\mathcal{H}$ $>$ $\mathcal{H}_c$: Purely subsonic accretion and outflow with vanishing \\
             &  speed at horizon and non-zero at spatial infinity  (blue curves within $-v_c < v < v_c$) \\
	IV  &	$\mathcal{H}$ $<$ $\mathcal{H}_c$	: Unphysical flow\\
        \br
\end{tabular}
\end{table}


\section{Black hole's mass accretion rate}
\label{sec:massrate}
 In this section, we estimate the mass accretion rate in general and variation of mass accretion rate is studied at different radii for a higher-dimensional RN BH. The mass accretion rate is referred to as the change in rate of the BH's mass which is a measure of the BH's mass per unit of time (represented by $\dot{\mathcal{M}}$). It is defined as the area times the flux of a BH at the horizon. In the usual four dimensions, the general expression of mass accretion rate is $\dot{\mathcal{M}}\,|_{r_{h}} = 4\pi \,r^2 \,T_t^r\,|_{r_{h}}$, where $T_t^r$ is the energy-momentum tensor of the perfect fluid. Since the energy of the dynamical system is conserved, we use the conservation equations defined in  equations (\ref{maineq1}) and (\ref{maineq2}) to obtain the following,
\begin{equation}
\label{ma1}
    r^{D-2}\,u\,(\rho + p)\,\Big(f(r) + u^2 \Big)^{1/2} = C_3\,,
\end{equation}
where $C_3$ is an arbitrary constant. Now, considering the equation of state $p = \omega \rho$ and the continuity equation given in equation (\ref{maineq1}), we get,
\begin{equation}
    \frac{d\rho}{\rho+p} + \frac{d\,u}{u} + (D-2) \frac{d\,r}{r}=0.
\end{equation}
Integrating the above equation yields,
\begin{equation}
\label{ma2}
    r^{D-2}\,u\,\exp \Bigg[ \int^{\rho}_{\rho_{\infty}} \frac{d\,\rho\,'}{\rho\,' + p(\rho\,')}\Bigg] = -C_4\,,
\end{equation}
where $C_4$ is the integration constant. The minus sign is assumed due to the negative value of $u$ and $\rho_{\infty}$ is the fluid density at infinity. Using equations (\ref{ma1}) and (\ref{ma2}), we obtain,
\begin{equation}
    C_5 = - \frac{C_3}{C_4} = (\rho + p)\,\Big(f(r) + u^2 \Big)^{1/2}\,\exp \Bigg[ -\int^{\rho}_{\rho_{\infty}} \frac{d\,\rho\,'}{\rho\,' + p(\rho\,')}\Bigg],
\end{equation}
where $C_5$ is an arbitrary constant. Considering the boundary at infinity, $C_5 \equiv \rho_{\infty} + p(\rho_{\infty}) = -\frac{C_3}{C_5}$, where $C_3 \equiv (\rho + p)\,u_t\,u\,r^{D-2} = -C_4\,(\rho_{\infty} + p(\rho_{\infty}) )$. Further making use of equations (\ref{maineq1}) and (\ref{ma1}), we get
\begin{equation}
    \frac{\rho + p}{\varrho}\,\Big(f(r) + u^2 \Big)^{1/2} = \frac{C_3}{C_1} \equiv C_6\,,
\end{equation}
where $C_6$ is an arbitrary constant, {\it i.e.}, $C_6 = \frac{(\rho_{\infty}\, +\, p(\rho_{\infty}))}{\varrho_{\infty}}$. From equation (\ref{ma1}), we obtain the new relation for BH's mass accretion rate
\begin{equation}
    \dot{\mathcal{M}} = -4\pi\,r^{D-2}\,u\,(\rho + p)\,\Big(f(r) + u^2 \Big)^{1/2} = -4\pi\,C_3\,,
\end{equation}
which  takes the  simple form
\begin{equation}
    \dot{\mathcal{M}} = 4\pi\,C_4\,(\rho_{\infty}\, +\, p(\rho_{\infty})).
\end{equation}
Using the boundary condition at infinity, the above equation is valid for any nature of fluid following the EoS $p=p(\rho)$. Thus, the BH mass accretion rate can be  defined as follows
\begin{equation}
    \label{massaccretion}
    \dot{\mathcal{M}} = 4\pi\,C_4\,(\rho\, +\, p(\rho))|_{r=r_h}\,,
\end{equation}
For isothermal fluid, the EoS, {\it i.e.} $p=\omega\,\rho$, leads to $(\rho + p) \equiv \rho(1+\omega)$. Thereafter from equation (\ref{ma2}), we obtain
\begin{equation}
    \label{madensity}
    \rho = \Bigg[ -\frac{C_4}{r^{D-2}\,u} \Bigg]^{1+\omega}\,,
\end{equation}
Using $\rho$ in equation (\ref{ma1}) we obtain a general equation given by
\begin{equation}
    \label{ma3}
    u^2 - \frac{C_3^2\,C_4^{-2(1+\omega)}}{(1+\omega)^2}\,r^{2\omega(D-2)}\,u^{2\omega}+f(r)=0\,.
\end{equation}
We obtain $u$ which is a function of  $\omega$ that determines the energy density of the fluid ($\rho(r)$) from equation (\ref{madensity}). The mass accretion rate ($\dot{\mathcal{M}}$) can be obtained from equation (\ref{massaccretion}).


\subsection{Case I: Utra-Stiff fluid ($\omega = 1$)}

Ultra-stiff fluid corresponds to $\omega=1$, and the radial component of the four-velocity can be determined from equation (\ref{ma3}) as
\begin{equation}
    u = \pm\, 2C_4^2 \sqrt{\frac{f(r)}{C_3^2}\,r^{2(D-2)}-4\,C_4^4}.
\end{equation}
Using equation (\ref{madensity}), we obtain the energy density as
\begin{equation}
\label{maomega1}
    \rho = \frac{C_3^2\,r^{2(D-2)}-4\,C_4^4}{4\,C_4^2\,r^{2(D-2)}\,f(r)}.
\end{equation}
Thus, the mass accretion rate for $\omega=1$ yields
\begin{equation}
    \dot{\mathcal{M}} = \frac{2\pi(C_3^2\,r^{2(D-2)}-4\,C_4^4)}{C_4\,r^{2(D-2)}\,f(r)}.
\end{equation}
In figure (\ref{fig:maa}), we plot the BH mass accretion rate ($\dot{\mathcal{M}}$) against the radius ($r$) for $\omega=1$ for even and odd dimensional BHs taking  $M = 1.5$ and $q = 1$. It is evident from the plot  that the mass accretion rate diminishes with an increase in the spacetime dimensions. The maximum mass accretion rate is 
\begin{itemize}
    \item $D=4$ : $\dot{\mathcal{M}}\, \approx \, 5.18967$ at $r\, \approx \, 4.51708$.
    \item $D=5$ : $\dot{\mathcal{M}}\, \approx \, 2.88525$ at $r\, \approx \, 3.98860$.
    \item $D=6$ : $\dot{\mathcal{M}}\, \approx \, 2.76515$ at $r\, \approx \, 3.33562$.
    \item $D=7$ : $\dot{\mathcal{M}}\, \approx \, 2.74341$ at $r\, \approx \, 2.88243$.
\end{itemize}
It is noted that the radius at which the mass accretion rate is maximum,  moves towards the singularity with the increase in spacetime dimension.


\subsection{Case II: Utra-Relativistic fluid ($\omega = \frac{1}{2}$)}
\begin{figure*}[t]
 \begin{subfigure}{0.49\textwidth}
     \includegraphics[width=\textwidth]{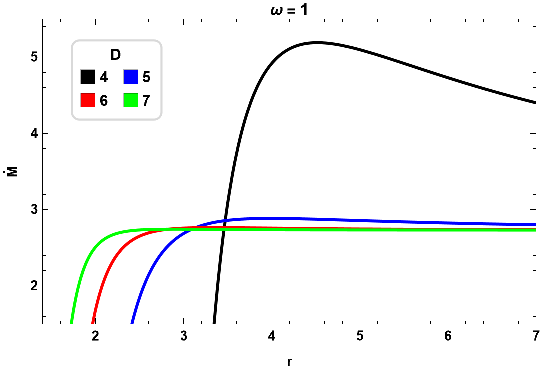}
     \caption{}
     \label{fig:maa}
 \end{subfigure}
 \hfill
 \begin{subfigure}{0.51\textwidth}
     \includegraphics[width=\textwidth]{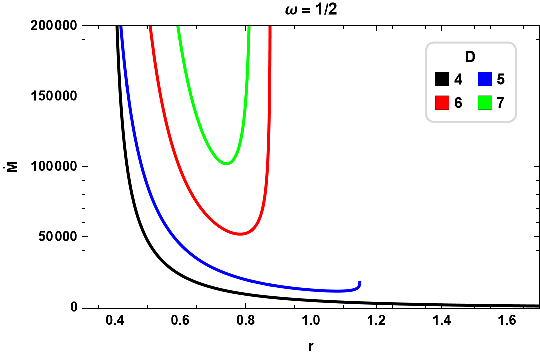}
     \caption{}
     \label{fig:mab}
 \end{subfigure}
 \medskip
 \begin{subfigure}{0.49\textwidth}
     \includegraphics[width=\textwidth]{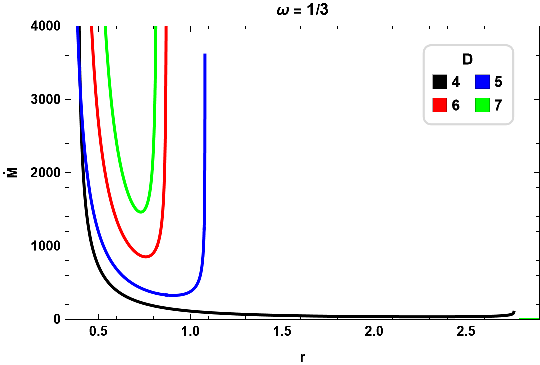}
     \caption{}
     \label{fig:mac}
 \end{subfigure}
 \hfill
 \begin{subfigure}{0.49\textwidth}
     \includegraphics[width=\textwidth]{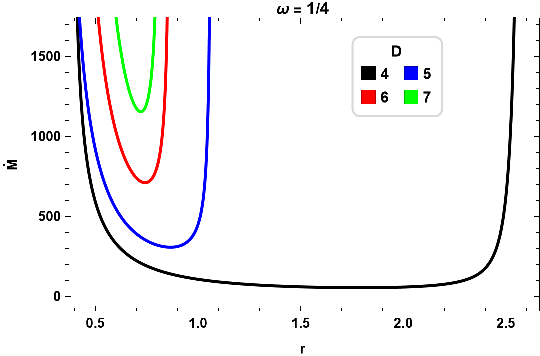}
     \caption{}
     \label{fig:mad}
 \end{subfigure}

 \caption{Variation of mass accretion rate against radius for a higher-dimensional RN BH with $M = 1.5$ and $q = 1$ for different values of state parameter and in different dimensions.}
 \label{fig:massrate}
\end{figure*}

For Utra-Relativistic fluid $\omega = \frac{1}{2}$, equation (\ref{ma3}) yields the radial component of the four-velocity as
\begin{equation}
    u = \frac{2C_3^2\, r^{D-2} + \sqrt{4C_3^4\,r^{2(D-2)} - 81 f(r)\,C_4^6}}{9C_4^3},
\end{equation}
Using equation (\ref{madensity}), energy density is determined
\begin{equation}
    \rho = 27 \Bigg[ \frac{C_4^4}{r^{D-2} \Big( 2C_3^2\,r^{D-2} + \sqrt{4C_3^4\,r^{2(D-2)} - 81 f(r)\,C_4^6}\Big)}\Bigg]^{\frac{3}{2}},
\end{equation}
The mass accretion rate for $\omega=\frac{1}{2}$ becomes
\begin{equation}
    \dot{\mathcal{M}} = 162\pi \, C_4\Bigg[ \frac{C_4^4}{r^{D-2} \Big( 2C_3^2\,r^{D-2} + \sqrt{4C_3^4\,r^{2(D-2)} - 81 f(r)\,C_4^6}\Big)}\Bigg]^{\frac{3}{2}}.
\end{equation}
In figure (\ref{fig:mab}), we plot the mass accretion rate against the radius for ultra-relativistic fluid ($\omega = \frac{1}{2}$) for even and odd dimensional BHs with parameters $M = 1.5$ and $q = 1$. We note that the mass accretion rate is the maximum near the two horizons, {\it i.e.} event horizon and the Cauchy horizon of the RN BH. The maximum mass accretion rate at the horizons are tabulated in table \ref{tab:ma1by2}. It is evident that for $D=4$, the mass accretion rate is maximum only at the Cauchy horizon. Interestingly, the value of the maximum mass accretion rate increases at both the horizons with the increase in spacetime dimension.

 \begin{table}[h] 
    \centering
\caption{\label{tab:ma1by2}}
Values of maximum accretion rate at the event and Cauchy horizon for $D = 4,\,5,\,6,\,7$ with suitable parameters $M = 1.5$, $q = 1$, $\omega = \frac{1}{2}$.
\begin{tabular}{c c c c c}
	D	&	$r_{ch}$	&	$(\dot{\mathcal{M}})_{r_{ch}}$	&	$r_h$	&	$(\dot{\mathcal{M}})_{r_{h}}$	  	\\
	\hline
	4	&	0.382	&	1.9 $\times$ $10^7$	&	 $-$       &	$-$		\\
	5	&	0.342	&	7.4 $\times$ $10^8$	&	1.15     &	17499	\\
	6	&	0.386	&	3.5 $\times$ $10^9$ &	0.875    &	301623		\\
	7	&	0.437	&	8.1 $\times$ $10^9$ &	0.82	 &	1.2 $\times$ $10^6$		\\	
\end{tabular}
\end{table}


\subsection{Case III: Radiation fluid ($\omega = \frac{1}{3}$)}
For radiation fluid ($\omega=\frac{1}{3}$), the radial component of the four-velocity following equation (\ref{ma3}) is given by
\begin{eqnarray}
    u = \frac{1}{8} \Bigg[ \frac{\Big( C_4^2 (-32C_4^4\,f(r) + \sqrt{1024f(r)^2\,C_4^8 - 27C_3^6\,r^{2(D-2)}}) \Big)^{\frac{1}{3}}}{C_4^2}   \nonumber \\
    + \frac{3C_3^2\,r^{\frac{2(D-2)}{3}}}{C_4^{\frac{2}{3}} \Big( C_4^2 (-32C_4^4\,f(r) + \sqrt{1024f(r)^2\,C_4^8 - 27C_3^6\,r^{2(D-2)}}) \Big)^{\frac{1}{3}}} \Bigg]^{\frac{2}{3}},
\end{eqnarray}
The energy density using equation (\ref{madensity})  is obtained as
\begin{eqnarray}
    \rho = 16 \Big(\frac{C_4}{r^{D-2}}\Big)^{\frac{4}{3}} \Bigg[ \frac{\Big( C_4^2 (-32C_4^4\,f(r) + \sqrt{1024f(r)^2\,C_4^8 - 27C_3^6\,r^{2(D-2)}}) \Big)^{\frac{1}{3}}}{C_4^2}   \nonumber \\
    + \frac{3C_3^2\,r^{\frac{2(D-2)}{3}}}{C_4^{\frac{2}{3}} \Big( C_4^2 (-32C_4^4\,f(r) + \sqrt{1024f(r)^2\,C_4^8 - 27C_3^6\,r^{2(D-2)}}) \Big)^{\frac{1}{3}}} \Bigg]^{-2},
\end{eqnarray}
The mass accretion rate is
\begin{eqnarray}
    \dot{\mathcal{M}} = \frac{256}{3}\pi\,C_4 \Bigg[ \frac{\Big( C_4^2 (-32C_4^4\,f(r) + \sqrt{1024f(r)^2\,C_4^8 - 27C_3^6\,r^{2(D-2)}}) \Big)^{\frac{1}{3}}}{C_4^2}   \nonumber \\
    + \frac{3C_3^2\,r^{\frac{2(D-2)}{3}}}{C_4^{\frac{2}{3}} \Big( C_4^2 (-32C_4^4\,f(r) + \sqrt{1024f(r)^2\,C_4^8 - 27C_3^6\,r^{2(D-2)}}) \Big)^{\frac{1}{3}}} \Bigg]^{-2}.    
\end{eqnarray}

 \begin{table}[h] 
    \centering
\caption{\label{tab:ma1by3}}
Values of maximum accretion rate at the event and Cauchy horizon for $D = 4,\,5,\,6,\,7$ with suitable BH parameters $M = 1.5$, $q = 1$, $\omega = \frac{1}{3}$.
\begin{tabular}{c c c c c}
	D	&	$r_{ch}$	&	$(\dot{\mathcal{M}})_{r_{ch}}$	&	$r_h$	&	$(\dot{\mathcal{M}})_{r_{h}}$	  	\\
	\hline
	4	&	0.382	&	258667	            &	 2.760   &	99		\\
	5	&	0.342	&	831772             	&	1.080    &	3605	\\
	6	&	0.386	&	2.8 $\times$ $10^6$ &	0.871    &	17167		\\
	7	&	0.437	&	4.7 $\times$ $10^6$ &	0.819	 &	41444		\\	
\end{tabular}
\end{table}

In figure (\ref{fig:mac}), the mass accretion rate is plotted against the radius for radiative fluid ($\omega = \frac{1}{3}$) for even and odd dimensional BHs with parameters $M = 1.5$, $q = 1$. The mass accretion rate is maximum at the Cauchy and event horizon and minimum between the two horizons. The maximum values of the mass accretion rates and the approximate values of the horizons as mentioned earlier are tabulated in table \ref{tab:ma1by3}. We note that the maximum mass accretion rate at the horizons increases as the spacetime dimension increases.


\subsection{Case IV: Sub-Relativistic fluid ($\omega = \frac{1}{4}$)}
For sub-relativistic fluid ($\omega = \frac{1}{4}$), the radial component of the four-velocity, the energy density and the mass accretion rate are derived using the equations (\ref{ma3}), (\ref{madensity}) and (\ref{massaccretion}), respectively. The radial variation of the mass accretion rate is plotted in figure (\ref{fig:mad}) for even and odd dimensional BHs for the parameters $M = 1.5$, $q = 1$. The maximum mass accretion rates along with the position of the said horizons are tabulated in table \ref{tab:ma1by4}. We note that the mass accretion rate is maximum at the Cauchy and event horizons which however attains a minimum value between the horizons. For $D=5$,  the mass accretion rate is found minimal at the Cauchy horizon, which, however, at the event horizon attains a maximum. We plot for various dimensions and found that after $D=6$, the accretion rate at both the horizons decreases with the increase in spacetime dimension. \\
 \begin{table}[t] 
    \centering
\caption{\label{tab:ma1by4}}
Values of maximum accretion rate at the event and Cauchy horizon for $D = 4,\,5,\,6,\,7$ with suitable BH parameters $M = 1.5$, $q = 1$, $\omega = \frac{1}{4}$.
\begin{tabular}{c c c c c}
	D	&	$r_{ch}$	&	$(\dot{\mathcal{M}})_{r_{ch}}$	&	$r_h$	&	$(\dot{\mathcal{M}})_{r_{h}}$	  	\\
	\hline
	4	&	0.382	&	1.9 $\times$ $10^{13}$	&	2.617    &	8.8 $\times$ $10^{7}$\\
	5	&	0.342	&	588704             	&	1.075    &	1.2 $\times$ $10^{14}$	\\
	6	&	0.386	&	6.2 $\times$ $10^{12}$ &	0.870    &	2.2 $\times$ $10^{13}$\\
	7	&	0.437	&	4.1 $\times$ $10^{12}$ &	0.819	 &	1.1 $\times$ $10^{13}$\\	
\end{tabular}
\end{table}
The energy released in the process of accretion can also be obtained by studying the maximum radiation output or maximum luminosity, which can be expressed as a function of mass accretion rate, {\it i.e.}, $L = \eta \, \dot{\mathcal{M}}\, c^2$ \cite{mat23,st83}, where $\eta$ is the efficiency of the radiative transfer. Therefore, the energy released as radiation is proportional to the maximum mass accretion rate. However, we note that the radiative efficiency is low for spherically symmetric accretion. In spherical accretion, most of the gravitational energy of the accreting material is not radiated away but falls into the BH, which leads to inefficiencies in converting mass into radiation. The efficiency of the radiative transfer for spherical accretion ranges from $10^{-5}\,-\,10^{-3}$. Therefore, the maximum luminosity decreases with an increase in the spacetime dimensions for $\omega = 1$ and is maximum for a four-dimensional RN BH. For $\omega = \frac{1}{2},\,\frac{1}{3}$, the maximum luminisity increases with the spacetime dimensions. For $\omega=\frac{1}{4}$, the maximum luminosity differs for different spacetime dimensions. 

\begin{figure*}[ht]
\centering
\begin{tabular}{cc}
  \includegraphics[width=0.45\textwidth,height= 0.35\textwidth]{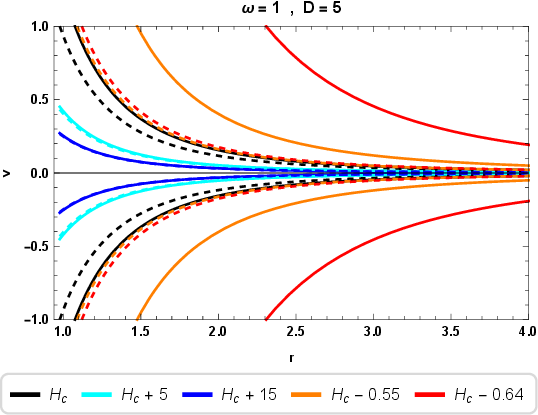} &   \includegraphics[width=0.45\textwidth,height= 0.35\textwidth]{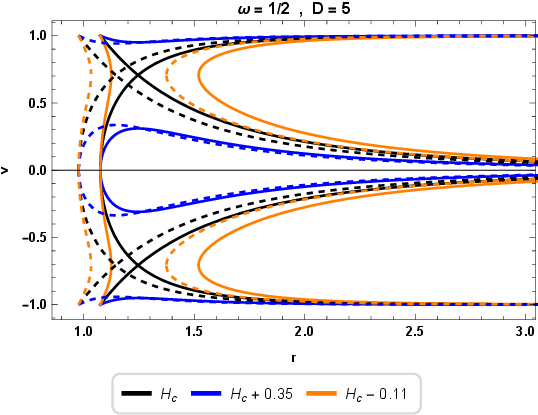}\\[6pt]
  \includegraphics[width=0.45\textwidth,height= 0.35\textwidth]{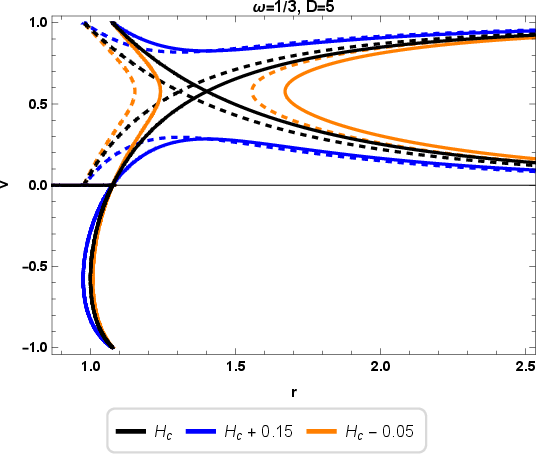} &   \includegraphics[width=0.45\textwidth,height= 0.35\textwidth]{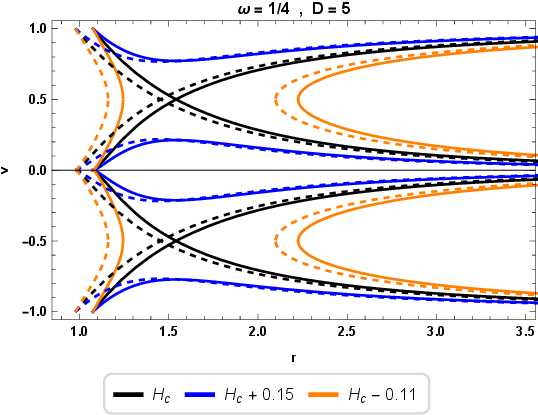}\\[6pt]
\end{tabular} 
\caption{Contour plots of  Hamiltonian $\mathcal{H}$   with different values of state parameter ($\omega$) in the RN BH for the parameter $M = 1.5$ and $D=5$. Solid curves correspond to $q = 1$. The dashed curve corresponds to $q = 1.5$.  } 
 	\label{fig:charge}
\end{figure*}


\section{Accretion of fluid with different charge parameters}
\label{sec:6}
In this section, we study the effect of charge on the accretion phenomena of  RN BH varying the charge of the fluid. The Charge-to-Mass limit for the accretion in the RN BH can be obtained from equation (\ref{ehradius}) which is,
\begin{equation}
    \label{CMratio}
	\frac{4M^2}{(D-2)^2} \geq \frac{2q^2}{(D-2)(D-3)}\ \
	\Rightarrow \ \ \frac{q^2}{M^2} \geq \frac{2(D-3)}{(D-2)}
\end{equation}  
It is found that the Charge-to-Mass ratio increases with an increase in dimensions. For a BH with mass parameter $M=1.5$, we obtain $q \leq 1.5$ as the limit for a 4-dimensional BH. As the dimension increases, we found that the limiting value increases. For a higher-dimensional RN BH with mass parameter $M=1.5$, the limiting value for the charge parameter is estimated numerically, these are 1.73205, 1.83712 and 1.89737 for $D=$ 5, 6 and 7, respectively. We have taken  $q = 0$ (Schwarzschild BH), $q = 1$ and $q = 1.5$ with a BH mass parameter, $M = 1.5$ and contours are drawn in figure (\ref{fig:charge}) for the Hamiltonian given in equation (\ref{finalhamiltonian}) for different values of the EoS state parameter ($\omega$). Critical values of the radius, the radial component of the four-velocity and the Hamiltonian are also tabulated in table \ref{tab:ch1} for different $\omega$ for $D=5$. It is evident from the figure (\ref{fig:charge}) and table \ref{tab:ch1} that the critical radius decreases as the charge increases, however, the critical Hamiltonian increases with an increase in the charge parameter for all values of $\omega$ considered. It is crucial to point out that the accretion profile obtained above for different state parameters $\omega$ is similar to the isothermal fluid case and a change in the charge parameter does not affect the flow behaviour.

\begin{table}[h] 
\centering
\caption{\label{tab:ch1}}
Values of $r_c$, $v_c$, and $H_c$ at the sonic point for $\omega  = 1,\,\frac{1}{2},\,\frac{1}{3},\,\frac{1}{4}$ for BH charge parameters $q=0,\,1,\,1.5$ with suitable BH mass parameter $M = 1.5$ and dimension $D=5$.
\begin{tabular}{c c c c c c}
        \br
	$\omega$  & q	 &	$r_h$     &	$r_c$     &	$v_c$	    &	$H_c$	   \\
	\mr
 		     & 0	&	1.12838  &	1.12838  &	0.999998   &	0.484473   \\
	1	      & 1	 &	1.07537	  &	1.07537	  &	  1		    &	0.64663    \\
 		     & 1.5	&	0.977205 &	0.977205 &	  1	       &	1.14838   \\
        \mr
 		     & 0	&	1.12838  &	1.30294  &	  0.707107 &	0.452092  \\
  $\frac{1}{2}$	& 1	   &  1.075370  &	1.24605	&	 0.707105 &    0.502195   \\
 		     & 1.5	&	0.977205 &	1.14513  &	  0.707105 &	0.604224  \\
        \mr
 		     & 0	&	1.12838  &	1.45973  &	  0.577348 &	0.48348   \\
  $\frac{1}{3}$	& 1	   &  1.075370  &	1.40048	&	 0.577351 &	   0.510782   \\
 		     & 1.5	&	0.977205 &	1.30631  &	  0.577347 &	0.55864   \\
        \mr
 		     & 0	&	1.12838  &	1.59577  &	  0.5      &	0.517597  \\
  $\frac{1}{4}$	& 1	   &  1.075370  &  1.54139	&	 0.50002  &    0.535409   \\	
 		     & 1.5	&	0.977205 &	1.45444  &	  0.50002  &	0.564272  \\
        \br
\end{tabular}
\end{table}

\begin{figure*}[h]
\centering
\begin{tabular}{cc}
  \includegraphics[width=0.45\textwidth,height= 0.3\textwidth]{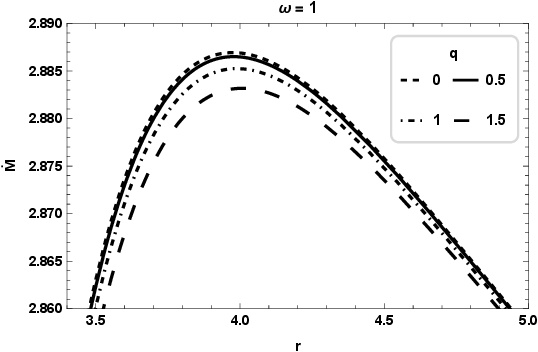} &   \includegraphics[width=0.45\textwidth,height= 0.3\textwidth]{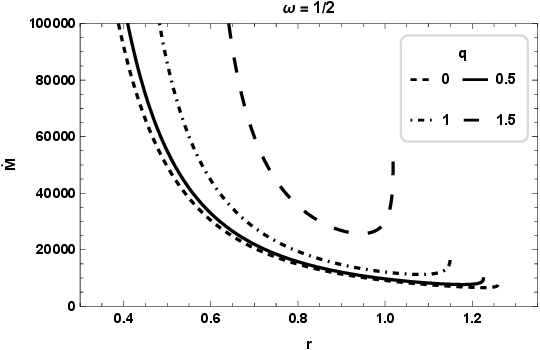}\\[6pt]
  \includegraphics[width=0.45\textwidth,height= 0.3\textwidth]{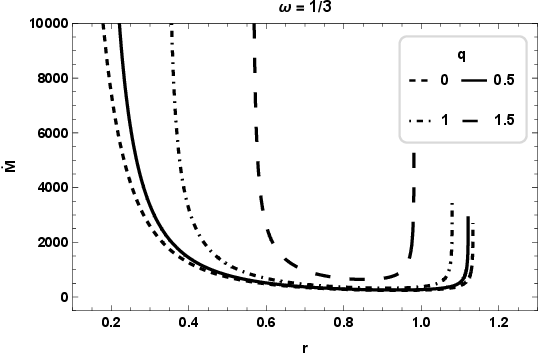} &   \includegraphics[width=0.45\textwidth,height= 0.3\textwidth]{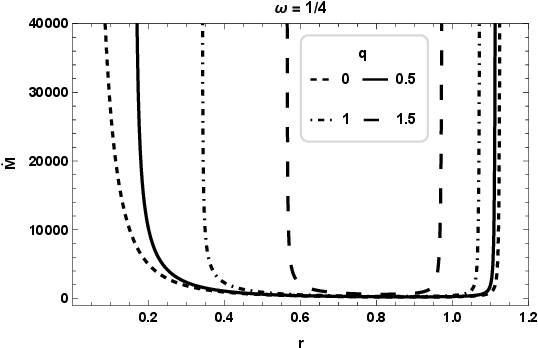}\\[6pt]
\end{tabular} 
\caption{Variation of mass accretion rate against radius for a higher-dimensional RN BH with $M = 1.5$ and $D = 5$ for different values of state parameter and different values of charge parameter ($q$)} 
 	\label{fig:chargerate}
\end{figure*}

{\it Effect on mass accretion rate :} In figure (\ref{fig:chargerate}), mass accretion rate is plotted against radius with different values of the charge parameter $q = 0,\,0.5,\,1,\,1.5$ and for different values of state parameter $\omega = 1,\,\frac{1}{2},\,\frac{1}{3},\,\frac{1}{4}$ of a RN BH with mass parameter $M=1.5$ and dimension $D=5$. Here $q=0$ corresponds to a Schwarzschild BH.  We note that for $\omega=1$, the maximum value of the mass accretion rate decreases with an increase in the charge parameter. The maximum mass accretion rate for $\omega = 1$ for different charge parameters is approximately
 \begin{itemize}
     \item $\dot{\mathcal{M}} \approx $ 2.88694 at $r \approx $ 3.97465 for $q=0$ .
     \item $\dot{\mathcal{M}} \approx $ 2.88651 at $r \approx $ 3.97814 for $q=0.5$ .
     \item $\dot{\mathcal{M}} \approx $ 2.88525 at $r \approx $ 3.97860 for $q=1$ .
     \item $\dot{\mathcal{M}} \approx $ 2.88318 at $r \approx $ 4.00608 for $q=1.5$ .
 \end{itemize}
                                    
\begin{table}[h] 
\centering
\caption{\label{tab:ch2}}
Maximum mass accretion rate $(\dot{\mathcal{M}})$  for $\omega  = 1,\,\frac{1}{2},\,\frac{1}{3},\,\frac{1}{4}$ for BH charge parameters $q=0,\,0.5,\,1,\,1.5$ with suitable BH mass parameter $M = 1.5$ and dimension $D=5$.
\begin{tabular}{c c c c c c}
        \br
	$\omega$  & q	 &	$r_{ch}$  &	$(\dot{\mathcal{M}})_{r_{ch}}$	&	$r_h$	&	$(\dot{\mathcal{M}})_{r_{h}}$  \\
        \mr
 		     & 0   &	$-$  &	         $-$             &	  1.259  &	7717  \\
  $\frac{1}{2}$	& 0.5 &  0.165  &	1.4 $\times$ $10^{11}$	&	 1.226  &  9810  \\
                  & 1	  &  0.342  &	7.4 $\times$ $10^{8}$	&	 1.150  &  17499  \\
 		     & 1.5 &  0.564  &	 1 $\times$ $10^{7}$     &	  1.019  &	51880  \\
        \mr
 		     & 0   &	$-$  &	         $-$             &	  1.133  &	2708  \\
  $\frac{1}{3}$	& 0.5 &  0.165  &	1.8 $\times$ $10^{7}$	&	 1.121  &  2883  \\
                  & 1	  &  0.342  &	0.85 $\times$ $10^{5}$	&	 1.080  &  3605  \\
 		     & 1.5 &  0.564  &	 0.17 $\times$ $10^{5}$  &	  0.981  &	6408  \\
        \mr
 		     & 0   &	$-$  &	         $-$             &	  1.128  &	2.3 $\times$ $10^{15}$  \\
  $\frac{1}{4}$	& 0.5 &  0.165  &	2.6 $\times$ $10^{7}$	&	 1.116  &  2.9 $\times$ $10^{15}$  \\
                  & 1	  &  0.342  &	0.58 $\times$ $10^{5}$	&	 1.075  &  1.2 $\times$ $10^{14}$  \\
 		     & 1.5 &  0.564  &	 1.9 $\times$ $10^{14}$  &	  0.977  &	1.2 $\times$ $10^{15}$  \\
        \br
\end{tabular}
\end{table}
The maximum mass rate at both the horizons for $\omega = \frac{1}{2},\,\frac{1}{3},\,\frac{1}{4}$ with different charge parameters are tabulated in table \ref{tab:ch2}. Since there is only one solution to $f(r)=0$ for a Schwarzschild BH ($q=0$), only one horizon exists. However, for $q=0$, we note that the mass accretion rate is very high near the singularity along with the a maximum mass accretion rate at the event horizon. We also note that for $\omega = \frac{1}{2}$ and $\frac{1}{3}$, the maximum mass accretion rate decreases near the Cauchy horizon, whereas the maximum mass accretion rate increases near the event horizon with an increase in the charge parameter. However for $\omega = \frac{1}{4}$, the maximum mass accretion rate is different for different values of charge parameters. Therefore, from the analysis of the mass accretion rate, we conclude that for ultra-stiff fluid ($\omega = 1$) and sub-relativistic fluid ($\omega = \frac{1}{4}$), the maximum mass accretion rate is higher for a Schwarzschild BH than that of a BH with charge. However, for ultra-relativistic fluid ($\omega = \frac{1}{2}$) and radiation fluid ($\omega = \frac{1}{3}$), the maximum mass accretion rate is lower for a Schwarzschild BH than that of a BH with charge.                                   

 
 \section{Discussion}
\label{sec:7}
 In the paper, we study spherically symmetric accretion onto a higher-dimensional Reissner-Nordstr\"{o}m (RN) black hole (BH) for two different fluids, namely, (i) Isothermal fluid and (ii) Polytropic fluid and explore the role of extra dimensions in the flow behaviour and accretion rate of the fluid.  The steady states for the fluids onto RN BH are determined analytically using a Hamiltonian dynamical system with BH mass parameter $M = 1.5$ and charge parameter $q = 1$. For this, we consider the two conservation laws: (i) conservation of the number density of particles and (ii) conservation of energy-momentum principles for the adiabatic processes of the matter accretion. The fluid density near the higher-dimensional BHs is assumed to be sufficiently small to neglect the self-gravity effects \cite{eo15}. We found some interesting features of the flow behaviour of the fluids near the RN BH for the fluids which follow the isothermal and polytropic equation of states (EoS), respectively. For the isothermal fluid, we analyze the accretion process onto a higher-dimensional RN BH with different values of the state parameters ($\omega$), they are classified as: {\it ultra-stiff fluid } (USF)($\omega = 1$), {\it ultra-relativistic fluid } (URF)($\omega = \frac{1}{2}$), {\it radiation fluid}  (RF)($\omega = \frac{1}{3}$), and {\it sub-relativistic fluid } (SRF)($\omega = \frac{1}{4}$).  In the case of polytropic fluids with the polytropic index satisfying the inequality: $1 < \Gamma < 2$, we analyze the accretion process. In this case, a realistic scenario of accretion is obtained for the adiabatic index $\Gamma = \frac{5}{3}$ in higher dimensions.  The role of additional dimensions in the accretion process is also studied numerically.\\

The contour of the Hamiltonian is plotted in the phase space to study the accretion of isothermal fluid onto higher-dimensional RN BHs in the presence of USF, URF, RF, and SRF  in figures (\ref{fig:usf}), (\ref{fig:urf}), (\ref{fig:rf}), and (\ref{fig:srf}), respectively for $M=1.5$ and charge parameter, $q=1$. The contour plots are compared for two different spacetime dimensions in figures (\ref{fig:usf1}), (\ref{fig:urf1}), (\ref{fig:rf1}), and (\ref{fig:srf1}) for a given fluid which is listed below. The dimensional dependence of the critical radius ($r_c$) and the critical Hamiltonian ($H_c$) are plotted in figure (\ref{fig:drh}). Subsequently, the accretion behaviour of a polytropic fluid onto a higher-dimensional RN BH is drawn in figure (\ref{fig:poly1}) for adiabatic index $\Gamma = 5/3$. We note the following:
 \begin{itemize}
     \item The isotropic fluid flow shows transonic, subsonic and supersonic behaviour, however, polytropic fluid shows purely subsonic or purely supersonic flow behaviour in all dimensions. 
     
     \item It is evident that as the spacetime dimensions increase, the critical radius decreases initially to a minimum at $D=8$ (for $\omega = 1\,,\frac{1}{2}$) and $D=9$ (for $\omega = \frac{1}{3},\,\frac{1}{4}$), thereafter it increases. For a given state parameter $\omega=\frac{1}{2}$, the critical radius decreases with the increase in dimensions and attains a minimum at $D=8$ which is found to decrease further if the charge parameter is increased. Similar behaviour of variation of the critical radius is evident from figure (\ref{fig:drh}) but the minimum of critical radius for other $\omega$ is found different from $D=8$.   The critical Hamiltonian is found to increase with an increase in spacetime dimensions which attains a maximum at $D=10$ for all types of isotropic fluid independent of the BH charge parameter; thereafter it decreases. 

     \item For $\omega = 1$, the event horizon radius (equal to the critical radius)  decreases initially to a minimum value for a given dimension and then increases as the spacetime dimension increases.
\end{itemize}  
In higher dimensions, the surface gravity of a BH increases, which yields a faster emission rate \cite{xu20a}, and hence a BH in a higher dimension evaporates faster than that of a BH in the usual four-dimensions \cite{xu20b}. As a result, the BH horizon radius and the critical radius shrink with increasing spacetime dimensions. \\

   The radial variations of the mass accretion rate are plotted in figure (\ref{fig:massrate}) for known isotropic fluids namely,  USF, URF, RF, and SRF when $D=4,\,5,\,6,\,7$ with the BH mass parameter, $M=1.5$ and charge parameter, $q=1$. We note the following: 
\begin{itemize}
    \item The mass accretion rate is maximum at two different points, {\it i.e.} one at the Cauchy horizon and the other at the event horizon. However, for $\omega = 1$, we get only one maximum. For $\omega = \frac{1}{2}, \frac{1}{3}$, the mass accretion rate is always higher at the Cauchy horizon for all dimensions.

    \item The accretion rate decreases with an increase in the spacetime dimensions for $\omega = 1$. For $\omega = \frac{1}{2},\,\frac{1}{3}$, the mass accretion rate increases with the spacetime dimensions. For $\omega=\frac{1}{4}$, the mass accretion rate differs for different spacetime dimensions.

    \item The maximum luminosity is highest for a four-dimensional RN BH  when $\omega = 1$ and it decreases with an increase in the spacetime dimensions. For $\omega = \frac{1}{2},\,\frac{1}{3}$, the maximum luminosity increases with the spacetime dimensions and it is lowest in a four-dimensional RN BH. For $\omega=\frac{1}{4}$, the maximum luminosity differs for different spacetime dimensions.
\end{itemize}
The maximum mass accretion rate and consequently, the energy released as radiation from a higher dimensional BH is more than the usual four dimensions (Ref.\cite{xu20a}). \\

The limiting value of the charge is found to increase with dimension, which is evident from equation (\ref{CMratio}). The contours of the Hamiltonian for  BH charge parameter namely, $q=1$ and $q=1.5$ are plotted in figure (\ref{fig:charge}) for $\omega  = 1,\,\frac{1}{2},\,\frac{1}{3},\,\frac{1}{4}$ when $M=1.5$ and $D=5$. The mass accretion rate is also plotted in figure (\ref{fig:chargerate}) for different values of state parameter with BH charge parameters $q=0,\,1,\,1.5$ with suitable BH mass parameter $M = 1.5$ and dimension $D=5$. We note the following:
\begin{itemize}
    \item As the charge parameter ($q$) of the BH increases, the critical radius ($r_c$) decreases and the critical Hamiltonian ($H_c$) also increases for all types of state parameter in isotropic fluid with a fixed mass parameter ($M$) and spacetime dimension ($D$). The critical radius is maximum and the critical Hamiltonian is minimum for a Schwarzschild BH in higher dimensions. The critical velocity remains the same for all values of the BH charge parameter in different dimensions.

    \item  For $\omega=1$, the maximum mass accretion rate is highest for a higher-dimensional Schwarzschild BH and with an increase in charge the maximum mass accretion rate decreases. For $\omega=\frac{1}{2},\,\frac{1}{3}$ the maximum mass accretion rate is the lowest for a higher-dimensional Schwarzschild BH and the mass accretion rate increases with an increase in charge. For $\omega=\frac{1}{2}$, the mass accretion rate is maximum for a higher-dimensional Schwarzschild BH.
\end{itemize}

A realistic accretion scenario is probed for a higher-dimensional RN BH employing Hamiltonian formalism for polytropic fluid with $\Gamma = 5/3$. The mass accretion rate of a higher-dimensional RN BH for polytropic fluid with $1 < \Gamma<\frac{5}{3}$ is obtained in the Ref.\cite{sha16} without Hamiltonian formalism, this indicates the limitation of the Hamiltonian dynamics when applied for accretion process. In the case of accretion onto a higher-dimensional BH employing the Hamiltonian approach, we get solutions for $\Gamma=\frac{5}{3}$ for the mass parameter $M=1.5$, this may be the fine-tuning and variability issue that exists in the Hamiltonian dynamical systems. However, in this formalism, an interesting result is obtained for the first time yielding the variation of the critical radius against the spacetime dimensions, {\it i.e.} the critical radius attains a minimum value as the spacetime dimension increases, depending upon the choice of state parameter of isotropic fluid. The reason for this new behaviour is an interesting topic to be discussed elsewhere. We consider a static RN BH, it is equally important to apply the methodology in a rotating RN BH,  to find more physics of spinning astrophysical objects in future.


\section{Acknowledgements}
\small
BD is thankful to CSIR, New Delhi, for financial support. BCP acknowledge SERB-DST, Govt. of India for Research Grant Ref. No. CRG/000183/2021. The authors would like to thank the IUCAA Centre for Astronomy Research and Development (ICARD), NBU, for extending research facilities. The authors are thankful to the anonymous Referees for the illuminating suggestions that have significantly improved in presenting the manuscript in its current form.

\vspace{5cm}

{\bf Reference }
\vspace{0.5cm}


\end{document}